\documentclass[12pt]{article}
\usepackage{amsmath,amsfonts,epsfig,amssymb}

\newcommand{\boldX}{\mathbf{X}}

\newcommand{\boldW}{\mathbf{W}}
\newcommand{\boldD}{\mathbf{D}}
\newcommand{\boldS}{\mathbf{S}}
\newcommand{\boldK}{\mathbf{K}}
\newcommand{\boldU}{\mathbf{U}}

\newcommand{\bolda}{\mathbf{a}}

\newcommand{\er}{\mathbb{R}}

\newcommand{\dq}{\mbox{d}q}

\newcommand{\picturesAB}[3]{
\centerline{\raise 2mm \hbox{\raise #3 \hbox{(a)}}
\hspace*{-.3in}
\psfig{file=#1,height=#3}
\hspace*{.02in}
\raise 2mm \hbox{\raise #3 \hbox{(b)}}
\hspace*{-.3in}
\psfig{file=#2,height=#3}
}}
\newcommand{\picturesABC}[4]{
\centerline{\raise #4 \hbox{(a)}
\psfig{file=#1,height=#4}
\hspace*{.2in}
\raise #4 \hbox{(b)}
\psfig{file=#2,height=#4}
\hspace*{.2in}
\raise #4\hbox{(c)}
\psfig{file=#3,height=#4}
}}
\newcommand{\norm}[1]{\parallel\!\!#1\!\!\parallel}
\numberwithin{equation}{section}
\long\def\symbolfootnote[#1]#2{\begingroup%
\def\thefootnote{\fnsymbol{footnote}}\footnote[#1]{#2}\endgroup}

\newcommand{\appsection}[1]{\let\oldthesection\thesection
\renewcommand{\thesection}{Appendix \oldthesection}
\section{#1}\let\thesection\oldthesection}

\begin{document}

\title{Variable-free exploration
of stochastic models: \\ a gene regulatory network example}

\author{
Radek Erban\thanks{University of Oxford, Mathematical Institute,
24-29 St. Giles', Oxford, OX1 3LB, United Kingdom;
{\it e-mail:  erban@maths.ox.ac.uk}.}
 \and
Thomas A. Frewen\thanks{Princeton University,
Department of Chemical Engineering, PACM and Mathematics,
Engineering Quadrangle,
Olden Street, Princeton, NJ 08544, USA;
{\it e-mail: tfrewen@princeton.edu, yannis@princeton.edu}.}
 \and
Xiao Wang\thanks{
Department of Statistics and Operations Research,
Bioinformatics and Computational Biology Program,
University of North Carolina at Chapel Hill,
Chapel Hill, NC 27599;
{\it e-mail: xiaow@email.unc.edu, telston@amath.unc.edu }}
 \and
 Timothy C. Elston$^\ddagger$
 \and
  Ronald Coifman\thanks{
Department of Mathematics,
 Yale University,
 New Haven, CT 06520;
{\it e-mail: Ronald.Coifman@yale.edu}}
\and
Boaz Nadler\thanks{
Department of Computer Science and Applied Mathematics,
Weizmann Institute of Science,
 Rehovot 76100 , Israel;
{\it e-mail: boaz.nadler@weizmann.ac.il}}
 \and
 Ioannis G. Kevrekidis$^\dagger$
}

\date{\today}

\maketitle
\vskip -1cm

\begin{abstract}
\noindent
Finding coarse-grained, low-dimensional descriptions is an important task in the
analysis of complex, stochastic models of gene regulatory networks.
This task involves (a) identifying observables that
best describe the state of these complex systems and (b) characterizing the dynamics
of the observables.
In a previous paper \cite{Erban:2006:GRN},
we assumed that good observables were known {\em a priori},
and presented an equation-free approach
to approximate coarse-grained quantities (i.e, effective drift and diffusion
coefficients) that characterize the long-time behavior of the
observables.
Here we use diffusion maps \cite{Coifman:2005:GDT} to extract appropriate observables
(``reduction coordinates") in an {\em automated} fashion;
these involve the leading eigenvectors of a weighted Laplacian on a graph
constructed from network simulation data.
We present lifting and restriction procedures for
translating between physical variables
and these data-based observables.
These procedures allow us to perform equation-free coarse-grained, computations
characterizing the long-term dynamics
through the design and processing 
of short bursts of stochastic simulation initialized at appropriate
values of the data-based observables.
\end{abstract}

\newpage

\section{Introduction}
Gene regulatory networks are complex high-dimensional stochastic dynamical systems.
These systems are subject to large intrinsic
fluctuations that arise from the inherent random nature of the biochemical reactions
that constitute the network.
Such features make realistic modeling of
genetic networks, based on exact representations of the Chemical Master Equation
(such as the Gillespie Stochastic Simulation Algorithm, SSA \cite{Gillespie:1977:ESS})
computationally expensive.
Recently there has been considerable work devoted to developing efficient
numerical algorithms for accelerating the stochastic
simulation of gene regulatory networks
\cite{Adalsteinsson:2004:BNS,Gillespie:2001:AAS,Erban:2006:GRN,Salis:2005:AHS}
and, more generally, of chemical reaction networks.
Many of these techniques are based on
time-scales separation and classify the biochemical reactions as ``slow'' or
``fast'' \cite{Rao:2003:SCK,Cao:2005:SCS,Haseltine:2002:ASC,E:2005:NSS,Chatterjee:2005:BDB}.
In this paper we combine such acceleration methods with recently developed
data-mining techniques (in particular,
diffusion maps \cite{Coifman:2005:GDT,Coifman:2005:GDT2,Nadler:2005:DMS}) capable
of identifying appropriate coarse-grained variables (``observables", ``reduction coordinates")
based on simulation data.
These observables are then used in the context of accelerating stochastic
gene regulatory network simulations; they guide the design, initialization,
and processing of the results of short bursts of full-scale SSA computation.
These bursts of SSA are used to numerically solve the (unavailable in closed
form) evolution equations for the observables; such so-called equation-free methods
\cite{Kevrekidis:2003:EFM} for studying stochastic models
have been successfully applied to complex systems arising in different contexts
\cite{Haataja:2004:AHD,Hummer:2003:CMD,Sriraman:2005:CND}.
In the context of gene regulatory networks -- but with {\em known}
observables -- equation free modeling has been illustrated in \cite{Erban:2006:GRN};
here we extend the approach to the more general class of problems where
appropriate observables are unknown {\it a priori}.

We describe the state of a gene regulatory network through a vector
\begin{equation}
\boldX
=
[X_1,X_2,X_3,\dots,X_N]
\label{statevector}
\end{equation}
where $X_i$ are the numbers of various protein molecules, RNA molecules
and genes in the system.
The behavior of the gene regulatory network is described by the time
evolution of the vector $\boldX(t)$.
For naturally occurring gene regulatory networks
the dimension $N$ of the vector $\boldX(t)$ is, in general,
moderately large, ranging from tens to hundreds of species.
However, the temporal evolution of the network over time scales of
interest can be often usefully described by a much smaller number
$n$ of coordinates.
For example, in \cite{Erban:2006:GRN}, we studied various models of
a genetic toggle switch with $N=2$, $N=4$ and $N=6$ components of
the vector $\boldX$; yet in all cases, the slow dynamics was
effectively one-dimensional, and a single linear combination of protein concentrations was
sufficient to describe the system, i.e. $n=1$.
In this paper we show how, for this genetic network system, good
coarse variables can be found by data-mining type methods based on
the diffusion map approach.

This paper is organized as follows:
We begin with a brief description of our model in Section
\ref{model}.
Section \ref{secEFsummary} quickly reviews the equation-free
approach for this type of bistable dynamics.
Given a low-dimensional set of observables, the main idea is to
locally estimate drift and diffusion coefficients of an unavailable
Fokker-Planck equation in these observables from short bursts of
appropriately initialized full stochastic simulations.
In Section \ref{secVFtheory} we show how to process the data
generated by stochastic simulations to obtain data-driven
observables through the construction of diffusion maps
\cite{Nadler:2005:DMS,Nadler:2005:DMS2}.
The leading eigenvectors of the weighted graph Laplacian defined on
a graph based on simulation data suggest appropriate ``automated''
reduction coordinates when these are not known {\em a priori}.
Such observables are then used to perform ``variable-free''
computations.
In Section \ref{secVarFree} we present lifting and restriction
procedures for translating between physical system variables
and the automated observables.
The bursts of stochastic simulation required for equation-free
numerics are designed (and processed)  based on these new
coordinates.
This combined ``variable-free equation-free" analysis appears
to be a promising approach for computing features of the long-time,
coarse-grained behavior of certain classes of complex stochastic
models (in particular, models of gene regulatory networks), as an
alternative to long, full SSA simulations.
The approach can, in principle, also be wrapped around different
types of full atomistic/stochastic simulators, beyond SSA, and in
particular accelerated SSA approaches such as implicit tau--leaping
\cite{Rathinam:2003:SSC} and multiscale or nested SSA \cite{Cao:2005:MSS,E:2005:NSS}.

\section{Model Description}

\label{model}

Our illustrative example is a two gene network in
which each protein represses the transcription of the other gene
(mutual repression).
This type of system has been engineered in {\it E. coli} and is often referred to
as a genetic toggle switch \cite{Gardner:2000:CGT,Hasty:2002:EGC}.
The advantage of this simple system is that it allows us to
test the accuracy of computational methods by direct
comparison with results from long-time
stochastic simulations.
More details about the model can be found in \cite{Kepler:2001:STR} and
in our previous paper \cite{Erban:2006:GRN}.
The system contains two genes with operators $O_1$ and $O_2$,
two proteins $P_1$ and $P_2$, and the corresponding dimers, i.e. $N=6$ in
(\ref{statevector}).
The production of $P_1$ ($P_2$) depends on the
chemical state of the upstream operator $O_1$ ($O_2$).
If $O_1$ is empty then $P_1$ is produced at the rate
$\gamma_1$ and if $O_1$ is occupied by a dimer of $P_2$, then protein $P_1$
is produced at a rate $\epsilon_1 < \gamma_1.$
Similarly, if $O_2$ is empty then $P_2$ is produced at the rate
$\gamma_2$ and if $O_2$ is occupied by a dimer of $P_1$, then protein $P_2$
is produced at a rate $\epsilon_2 < \gamma_2.$
Note that, for simplicity, transcription and translation
are described by a single rate constant.
The biochemical reactions  are (compare with \cite{Erban:2006:GRN})
\begin{equation}
\qquad{\mbox{ \raise 0.851 mm \hbox{$\emptyset$}}}
 \;
\mathop{\stackrel{\displaystyle\longrightarrow}\longleftarrow}^{\gamma_1 O_1 + \varepsilon_1 \overline{P_2P_2O_1}}_{\delta_1}
{\mbox{ \raise 0.851 mm \hbox{$P_1$}}}
\qquad\qquad\qquad\qquad\qquad
{\mbox{ \raise 0.851 mm \hbox{$\emptyset$}}}
 \;
\mathop{\stackrel{\displaystyle\longrightarrow}\longleftarrow}^{\gamma_2 O_2 + \varepsilon_2 \overline{P_1P_1O_2}}_{\delta_2}
{\mbox{ \raise 0.851 mm \hbox{$P_2$}}}
\label{prodP1P2}
\end{equation}
\begin{equation}
{\mbox{ \raise 0.851 mm \hbox{$P_1 + P_1$}}}
\;
\mathop{\stackrel{\displaystyle\longrightarrow}\longleftarrow}^{k_{1}}_{k_{-1}}
{\mbox{ \raise 0.851 mm \hbox{$\overline{P_1P_1}$}}}
\qquad\qquad\qquad\qquad\qquad
{\mbox{ \raise 0.851 mm \hbox{$P_2 + P_2$}}}
 \;
\mathop{\stackrel{\displaystyle\longrightarrow}\longleftarrow}^{k_{2}}_{k_{-2}}
{\mbox{ \raise 0.851 mm \hbox{$\overline{P_2P_2}$}}}
\label{prodP1P1P2P2}
\end{equation}
\begin{equation}
{\mbox{ \raise 0.851 mm \hbox{$\overline{P_2P_2}+O_1$}}}
\;
\mathop{\stackrel{\displaystyle\longrightarrow}\longleftarrow}^{k_{o1}}_{k_{-o1}}
{\mbox{ \raise 0.851 mm \hbox{$\overline{P_2P_2O_1}$}}}
\qquad\qquad\qquad\qquad
{\mbox{ \raise 0.851 mm \hbox{$\overline{P_1P_1}+O_2$}}}
 \;
\mathop{\stackrel{\displaystyle\longrightarrow}{\longleftarrow}}^{k_{o2}}_{k_{-o2}}
{\mbox{ \raise 0.851 mm \hbox{$\overline{P_1P_1O_2}$}}}
\label{onoffO1O2}
\end{equation}
where overbars denote complexes. Equations (\ref{prodP1P2}) describe production and degradation
of proteins $P_1$ and $P_2$.
Equations (\ref{prodP1P1P2P2})
are dimerization reactions and equations (\ref{onoffO1O2}) represent the
binding and dissociation of the dimer and DNA.

The state vector for our system is
\begin{equation}
\boldX = \left[
P_1,P_2,\overline{P_1P_1},\overline{P_2P_2},O_1,O_2
\right]
\label{statevectormutrep}
\end{equation}
where $P_1$ and $P_2$ are numbers of proteins,
$\overline{P_1P_1}$ and $\overline{P_2P_2}$ are numbers of
dimers and $O_1 \in \{ 0,1 \}$ and $O_2 \in \{ 0,1 \}$
are states of operators.
Assuming that we have just one
copy of Gene 1 and one copy of Gene 2 in the system, then
the values
of $O_1$ and $\overline{P_2P_2O_1}$, resp.
$O_2$ and $\overline{P_1P_1O_2}$, are related
by the conservation relations, namely
$$
\overline{P_2P_2O_1} = 1 - O_1,
\qquad
\mbox{resp.}
\qquad
\overline{P_1P_1O_2} = 1 - O_2.
$$
By virtue of (\ref{prodP1P2}), $O_1 = 1$ means that the first
protein is produced with rate $\gamma_1$, while
$O_1 = 0$ means that it is produced with
rate $\varepsilon_1 < \gamma_1$ (similarly for the second
protein).

Models such as the one defined by (\ref{prodP1P2}) --
(\ref{onoffO1O2}) can be validated experimentally, by comparing
their predictions with steady-state distributions of protein
abundances obtained through single cell fluorescence measurements of
intercellular variability in protein expression levels.

\section{Brief review of equation-free computations}
\label{secEFsummary}

Suppose we have a well-stirred mixture of $N$ chemically reacting
species; furthermore, assume that the evolution of the system can be
described in terms of $n <N$ slow variables (observables).
In the following we assume that $n=1$, and denote this variable $Q$.
The approach carries through
for the case of a relatively small number of slow variables as well.
The variable $Q$ might be the concentration of one of the chemical
species or some function of these concentrations (e.g. a linear
combination of some of them).
In Section
\ref{secVFtheory} we show how variable-free methods can be used to
suggest an appropriate $Q$.
Let ${\bf R}$  denote a vector of the remaining (fast, ``slaved"
system observables) which, together with $Q$ provide a basis for the
simulation space.
Our assumption implies that (possibly, after a short initial
transient) the evolution of the system can be approximately
described by the time-dependent probability density function
$f(q,t)$ for the slow variable $Q$ that evolves according to the
following {\it effective} Fokker-Planck equation
\cite{Risken:1989:FPE}:
\begin{equation}
\frac{\partial f}{\partial t}(q,t)
=
\frac{\partial}{\partial q}
\left(
-
V(q) f(q,t)
+
\frac{\partial}{\partial q}
[D(q) f(q,t)]
\right).
\label{FPE}
\end{equation}
If the effective drift $V(q)$ and the effective diffusion
coefficient $D(q)$ are explicitly known functions of $q$,
then (\ref{FPE}) can be used to compute interesting long-time
properties of the system (e.~g., the equilibrium distribution, 
transition times between metastable states).
Assuming that (\ref{FPE}) provides a good approximation 
\cite{Gardner:2000:CGT,Kepler:2001:STR}, and motivated by the formulas
\begin{eqnarray}
V(q)
&=&
\lim_{\Delta t \to 0}
\frac{ < Q(t+\Delta t) - q \, | \, Q(t)=q>}{\Delta t} \label{avgvel} \\
D(q)
&=&
\frac{1}{2} \,
\lim_{\Delta t \to 0}
\frac{ < [Q(t+\Delta t) - q]^2 \, | \, Q(t)=q>}{\Delta t}
\label{effdiff}
\end{eqnarray}
we used in
\cite{Haataja:2004:AHD,Hummer:2003:CMD,Kopelevich:2005:CGK,Sriraman:2005:CND}
the results of short $\delta$-function initialized simulation bursts
to estimate the average drift, $V$, and diffusion coefficient $D$.
Note that, in our context, the limit $\Delta t \to 0$ in
equations (\ref{avgvel}) and (\ref{effdiff})
should be interpreted as ``$\Delta t$ small, but not too small" , i.e.
the short bursts are short in the time scale of the slow
variable, yet long in comparison to the characteristic equilibration time of
the remaining system variables.

The steady solution of (\ref{FPE}) is proportional to
$\exp[-\beta\Phi(q)]$, where the effective
free energy $\Phi (q)$ is defined as
\begin{equation}
 \beta \Phi(q) =
- \int_0^q \frac{V(q^\prime)}{D(q^\prime)} \dq^\prime
+ \ln D(q) + \mbox{constant}.
\label{potentialPhi}
\end{equation}
Consequently, computing the effective free energy and the
equilibrium probability distribution can be accomplished without the need for
long-time stochastic simulations.

\noindent
A procedure for computationally estimating
$V(q)$ and $D(q)$ is as follows:

\leftskip 1cm
\bigskip

\noindent
{\bf (A)} Given $Q=q$, approximate the conditional
density $P({\bf r}|Q=q)$ for the fast variables $\bf R$.
Details of this preparatory step were given in \cite{Erban:2006:GRN}.

\noindent
{\bf (B)} Use $P({\bf r}|Q=q)$ from step (A) to determine
appropriate initial conditions for the short simulation bursts and run multiple realizations
for time $\Delta t$.
Use the results of these simulations and
the definitions (\ref{avgvel}) and (\ref{effdiff}) to estimate
the effective drift $V(q)$ and the effective diffusion coefficient $D(q)$.

\noindent
{\bf (C)} Repeat steps (A) and (B) for sufficiently many values of $Q$ and then compute
$\Phi(q)$ using formula (\ref{potentialPhi}) and numerical quadrature.

\leftskip 0cm
\bigskip

\noindent
Determining the accuracy of these estimates and, in particular, the
number of replica simulations required for a prescribed accuracy, is
the subject of current work.
An important feature of this algorithm is that it is trivially
parallelizable (different realizations of short simulations starting
at ``the same $q$" as well as realizations starting at different $q$
values can be run independently, on multiple processors).

\begin{figure}[ht]
\centerline{
\psfig{file=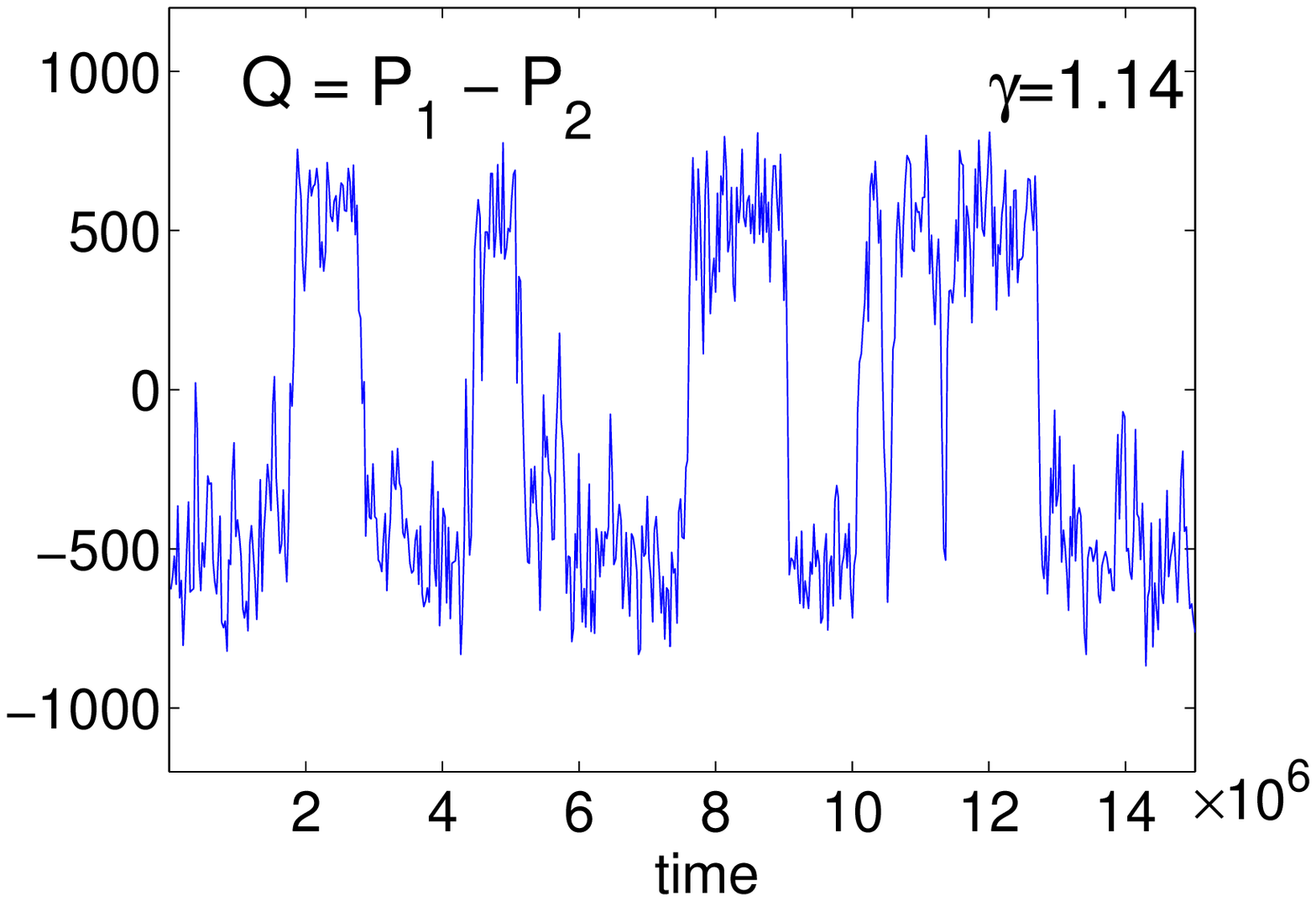,height=2.2in}
\quad
\psfig{file=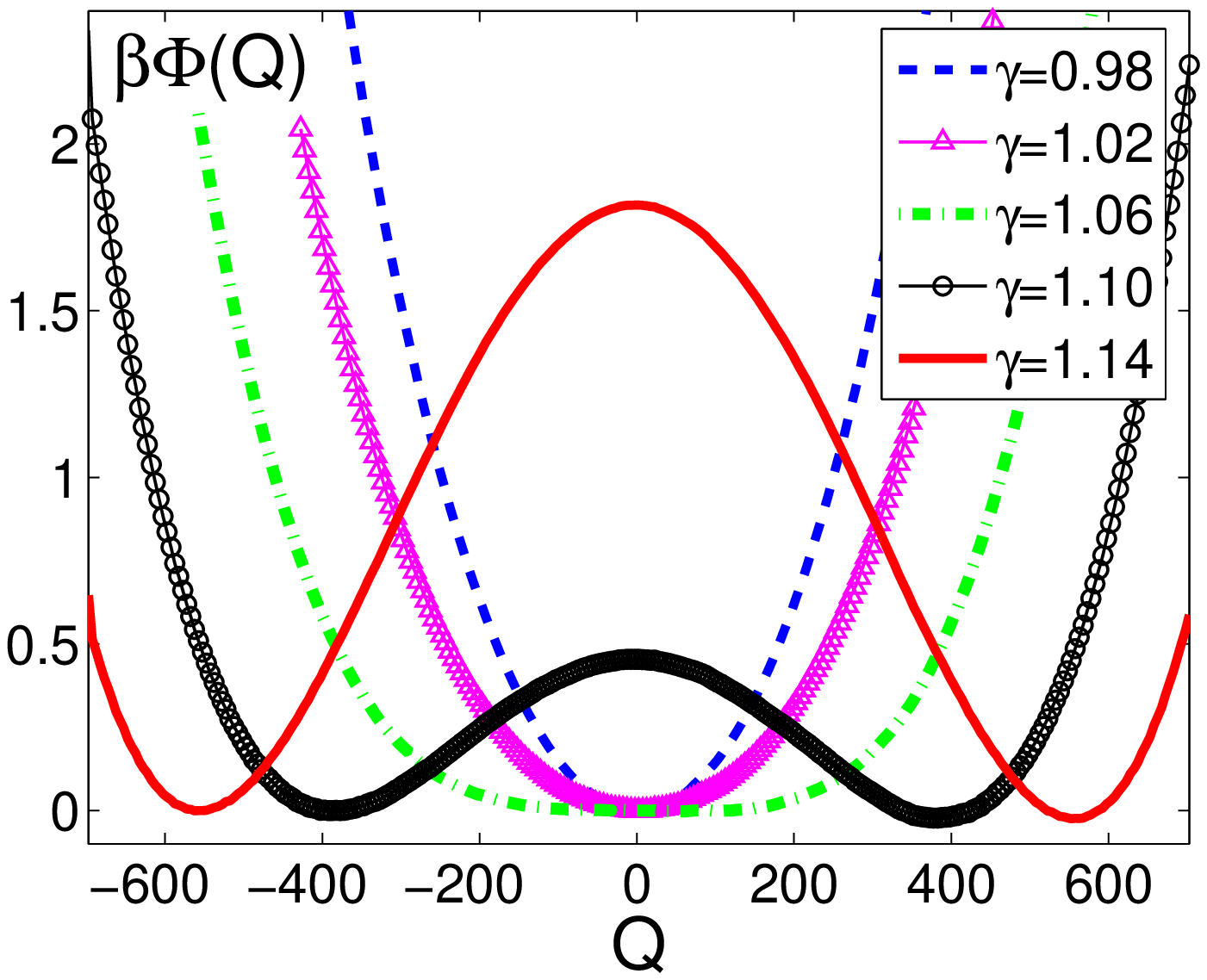,height=2.2in}
}
\smallskip
\centerline{
\psfig{file=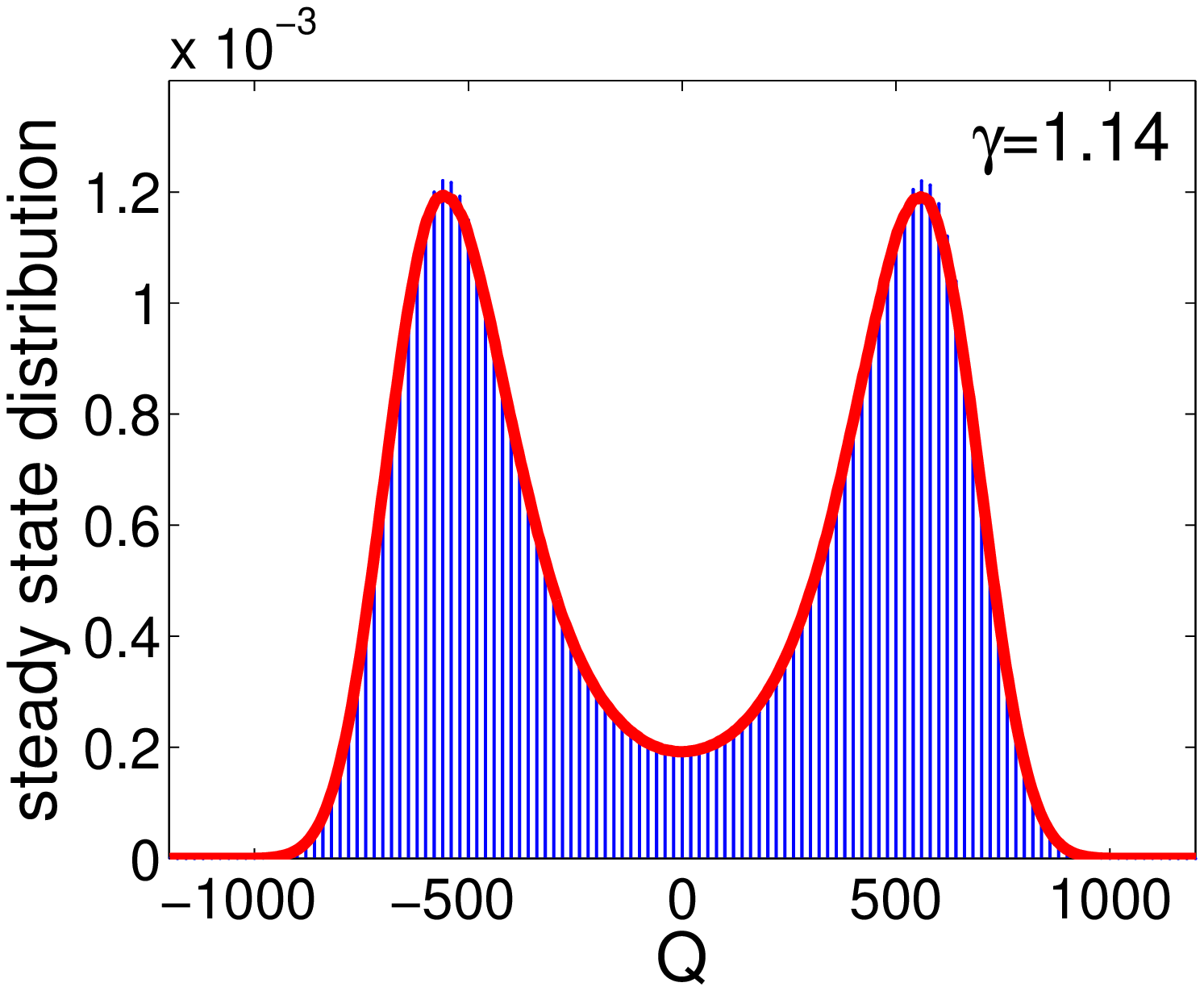,height=2.2in}
\quad
\psfig{file=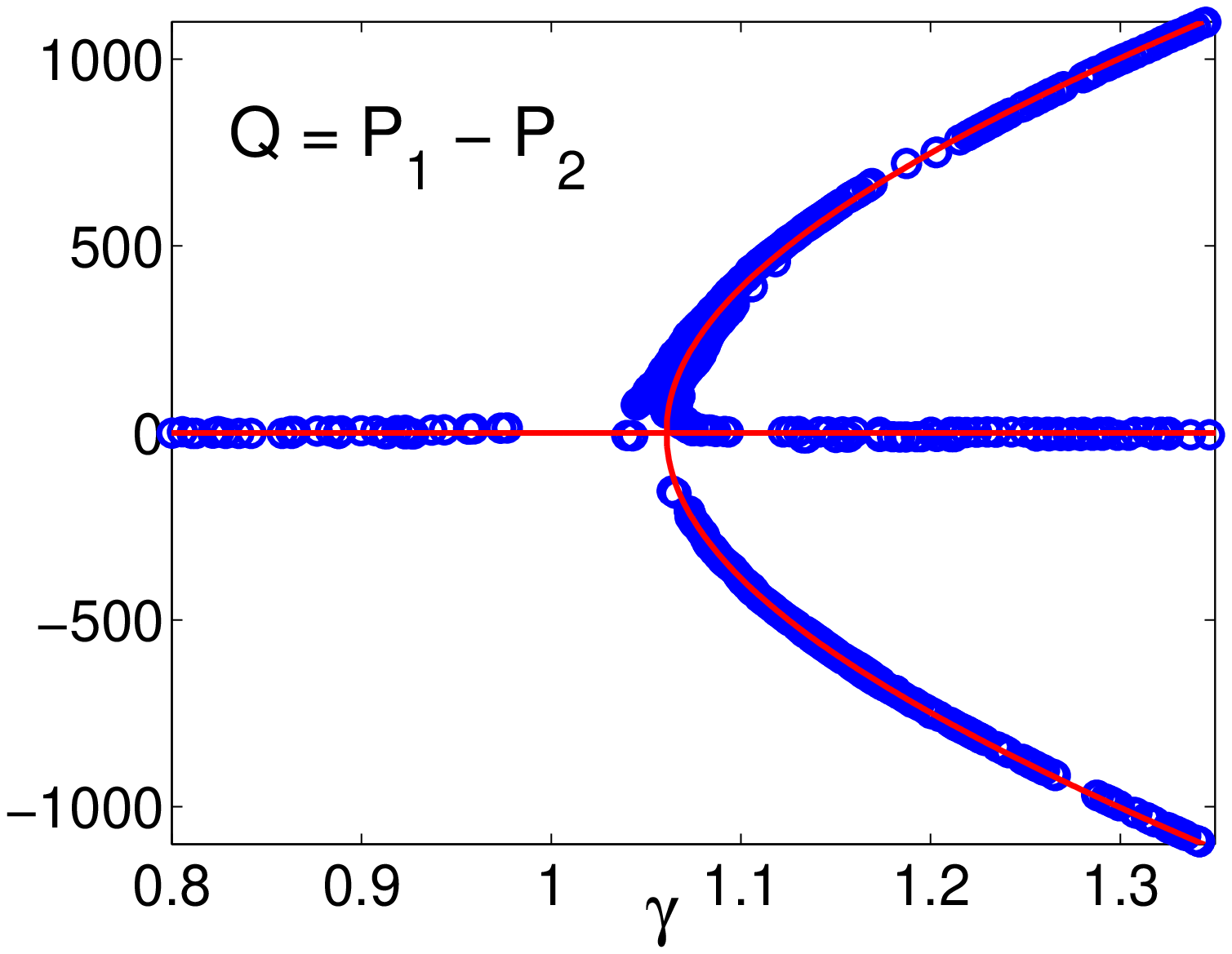,height=2.2in}
}
\caption{{\it Summary of equation-free results from {\rm \cite{Erban:2006:GRN}}.
To compute the figures, we used model $(\ref{prodP1P2})$ --
$(\ref{onoffO1O2})$ where equations $(\ref{prodP1P1P2P2})$ --
$(\ref{onoffO1O2})$ were assumed to be at quasi-equilibrium,
for parameter values see caption of 
Figure $5$ in {\rm \cite{Erban:2006:GRN}}.}}
\label{figSUMEF}
\end{figure}

A representative selection of equation-free results from our
previous paper \cite{Erban:2006:GRN}, for a stochastic model of a gene
regulatory network, is provided in Figure \ref{figSUMEF}. In \cite{Erban:2006:GRN} the (good)
observable $Q$ was assumed to be known {\em a priori}. The upper left panel
in Figure \ref{figSUMEF} shows a sample time series of $Q$, clearly
indicative of bistability, generated using the
stochastic model, while the upper right panel shows the effective free
energy $\beta\Phi$ computed using (\ref{potentialPhi}) as the parameter $\gamma \equiv \gamma_1 = \gamma_2$ is varied. The equation-free
steady state distribution of $Q$ obtained from this effective free
energy is in excellent agreement with histograms produced using
long-time simulation (lower left panel). Equation-free computation has
also been used \cite{Erban:2006:GRN} to 
compute ``stochastic bifurcation diagrams'' (an example is
shown in the bottom right
panel of Figure \ref{figSUMEF}) using an extension of deterministic bifurcation computation
\cite{Siettos:2003:CBD}. We believe this array of equation-free numerical
techniques holds promise for the acceleration of computer-assisted analysis of
gene-regulatory networks. We now extend this analysis to systems where the ``good'' observables are
unknown {\em a priori} by describing diffusion-map based variable-free methods.
\section{Variable-free methods}
\subsection{Theoretical framework}
\label{secVFtheory}

To find a good, low ($n$-)dimensional representation of the full
$N$-dimensional stochastic simulation data, we start by
exploring the phase-space of most likely configurations of the system through
extensive stochastic simulations; these configurations $\bf X$ (or a
representative sampling of them) at, say $M$ different times are
stored for processing.
From $M$ such recordings we obtain a set of $M$ vectors
$\bf{X}^{(1)},\ldots,\bf{X}^{(M)}$ in $\mathbb{R}^N$ which 
constitute the input to the diffusion map dimensionality reduction
approach we will now describe.
%
%
A crucial step for dimensionality reduction is the definition of a
meaningful {\em local } distance measure between configurations.
For continuous systems with equal noise strengths in all variables,
one may use the following pairwise similarity matrix
\begin{equation}
  \widetilde{W}_{ij} = \exp\left[-\left(\frac{\norm{{\bf X}^{(i)} - {\bf X}^{(j)}  }}{\sigma}\right)^2\right]
\label{weightmatrixtilda0}
\end{equation}
where $\norm{\cdot}$ is the standard Euclidean norm in
$\mathbb{R}^N$ and $\sigma$ is a characteristic scale for the
exponential kernel which quantifies the ``locality" of the
neighborhood in which the Euclidean distance is considered
(dynamically) meaningful \cite{Coifman:2005:GDT}.

For discrete chemical and biological reactions, as well as in other
systems where the components of the data vectors may be disparate
quantities varying over different orders of magnitude (possibly
including even Boolean variables), the simple Euclidean norm in
equation (\ref{weightmatrixtilda0}) with a single scaling factor
$\sigma$ equal for all components may, of course, not be
appropriate.
In this case, it is reasonable to consider different scalings for
the $N$ different components, using an $N$-dimensional weight vector
\begin{equation}
  {\mathbf a} = \left[a_1,a_2,\ldots,a_N\right]
  \label{weightvector}
\end{equation}
where $a_i>0$, for $i=1,\dots, N$, and define a {\em weighted} Euclidean norm
\begin{equation}
  \norm{{\bf X} }_{\mathbf a}^2 = \sum_{j=1}^N (a_j X_j)^2.
\end{equation}
This norm replaces the standard Euclidean norm in equation 
(\ref{weightmatrixtilda0}), where we may now choose $\sigma=1$,
since this scaling can be absorbed into the vector ${\bf a}$; thus
we replace (\ref{weightmatrixtilda0}) by
\begin{equation}
 \widetilde{W}_{ij} 
 = 
 \exp\left[- \norm{{\bf X}^{(i)} - {\bf X}^{(j)}  }_{\mathbf a}^2 \right].
\label{weightmatrixnoeucl}
\end{equation}
The elements of the matrix $\widetilde{\mathbf W}$ are all less than or
equal to one. Nearby points have $\widetilde{W}_{ij}$ close to one,
whereas distant points have $\widetilde{W}_{ij}$ close to zero.
In the
diffusion map approach, given $\alpha\in[0,1]$ (the choice of this
parameter value is discussed later), we define the matrix ${\bf W}$
by
\begin{equation}
  W_{ij} = \left(\sum_{k=1}^M \widetilde{W}_{ik}\right)^{-\alpha}
  \left(\sum_{k=1}^M \widetilde{W}_{jk}\right)^{-\alpha} \widetilde{W}_{ij}
\label{weightmatrix}
\end{equation}
Next, we define a diagonal $M\times M$ normalization matrix $\bf D$
whose values are given by
\begin{equation}
  D_{ii} = \sum_{k=1}^M W_{ik}
\label{diagmatrix}
\end{equation}
Finally we compute the eigenvalues and right eigenvectors of the
matrix
\begin{equation}
  {\bf K} = {\bf D}^{-1} {\bf W}.
\end{equation}
In this paper we will mainly work with the parameter $\alpha=0$.
However, in other applications different values of $\alpha$
may be more suitable (see \ref{appDMAP}).
As discussed in
 \cite{Nadler:2005:DMS,Belkin:2003:EDR,Nadler:2005:DMS2},
 if there
exists a {\em spectral gap} among the eigenvalues of this matrix,
then the leading eigenvectors may be used as a basis for a low
dimensional representation of the data (see \ref{appDMAP}).
To compute these eigenvectors, we can make use of the fact that
\begin{equation}
  {\bf K } = {\bf D}^{-1/2} {\bf S} {\bf D}^{1/2}\quad\mbox{  where  }
  {\bf S}  = {\bf D}^{-1/2} {\bf W} {\bf D}^{-1/2}
\label{symmatrix}
\end{equation}
is a symmetric matrix.
Hence, $\bf K$ and $\bf S$ are adjoint and
they have the same eigenvalues. Since $\bf S$ is symmetric, it is
diagonalizable with a set of $M$ eigenvalues
\begin{equation}
  \lambda_0 \geq \lambda_1 \geq \ldots \geq \lambda_{M-1}
\end{equation}
whose eigenvectors ${\bf U}_j$, $j=1,\ldots,M$ form an orthonormal
basis of $\mathbb{R}^M$.
The right eigenvectors of $\bf K$
are given by
\begin{equation}
  {\bf V}_j = {\bf D}^{-1/2} {\bf U}_j.
\label{eigenK}
\end{equation}
Since $\bf K$ is a Markov matrix, all its eigenvalues are smaller
than or equal to one in absolute value.
Moreover, if the parameter
$\sigma$ in (\ref{weightmatrixtilda0})
is large enough (and, thus, the norm vector in (\ref{weightmatrixnoeucl}) is ``small enough''), all points are (numerically) connected and
the largest eigenvalue $\lambda_0=1$ has multiplicity one with
corresponding eigenvector
\begin{equation}
  {\bf V}_0 = \left[1,1,\ldots,1\right].
\end{equation}
We define the $n$-dimensional representation of $N$-dimensional
state vectors by the following {\em diffusion map}
\begin{equation}
  \Psi_n : {\bf X^{(i)}} \to
  \left[V^{(i)}_1,V^{(i)}_2,\ldots,V^{(i)}_n\right];
\label{lowdimenrepresentation}
\end{equation}
that is, the point ${\bf X}^{(i)}$ is mapped to a vector containing
the $i$-th coordinate of each of the first $n$ leading eigenvectors
of the matrix ${\bf K}$.
This mapping $\Psi_n:\mathbb{R}^N\to \mathbb{R}^n$ is defined
{\em only} at the $M$ recorded state vectors.
We will show later that it can be extended to nearby points in the
$N$-dimensional phase space, without full re-computation of a new matrix
and its eigenvectors.
In \ref{appDMAP} we provide a theoretical justification for this
method as a dynamically useful dimensionality reduction step.

\subsection{Computation of data-based observables}

\label{secVFnumer}

We replaced (\ref{weightmatrixtilda0}) by (\ref{weightmatrixnoeucl})
where the weight vector (\ref{weightvector}) needs to be further specified.
Two natural choices for the values of components of the weight vector
$\bolda = [a_1,a_2, \dots, a_N]$ immediately arise.
One option is to regard the absolute values of the components of the state vector
$\boldX$ as of ``equal importance", i.e.
\begin{equation}
a_k = \omega,
\qquad
\mbox{for} \quad k=1, 2, \dots, N,
\label{veca1}
\end{equation}
where $\omega$ is a single method parameter; this is identical to
the use of a single $\sigma$ in equation \ref{weightmatrixtilda0},
namely $\sigma = \omega^{-1}.$

The above approach uses the Euclidean distance between data vectors 
as the basis for graph Laplacian construction and eigenanalysis.
In our case, the components of these vectors are concentrations 
of different species (e.g. integer numbers of protein molecules, 
each with its own range over the data set).
Moreover, the data vectors contain integers (0 and 1) representing
states of Boolean operators.
This motivates a second natural choice of the weight vector
$\bolda = [a_1,a_2, \dots, a_N]$. 
We rescale the state vector $X$ to span the
symmetrical domain (cube) in $N$-dimensional space, i.e.
\begin{equation}
a_k = \frac{\widetilde\omega}
{\displaystyle \max_i X_k^{(i)} - \displaystyle \min_i X_k^{(i)}},
\qquad
\mbox{for} \quad k=1, 2, \dots, N,
\label{veca2}
\end{equation}
where the maximum and minimum values are computed over all $i=1, \dots, M.$
Formula (\ref{veca2}) implies that components of the
vector $\boldX^{(i)}-\boldX^{(j)}$,
$i,j = 1, \dots, M,$ satisfy
$$
X_k^{(i)}-X_k^{(j)} \in \left[-\widetilde\omega,\widetilde\omega \right]
\qquad
\mbox{for} \; k=1, \dots, N, \; i,j = 1, \dots, M.
$$
The difference between (\ref{veca1}) and (\ref{veca2})
is that the first formula implicitly assumes that
the fluctuations in different components of the state vector
$\boldX$ are equally important, i.e. the absolute values
of fluctuations are important.
Formula (\ref{veca2}) on the other hand implies that
{\it relative} changes (compared to the maximal observed
change) in each component are more representative than the absolute
values of the changes.
We will see below that (\ref{veca1}) appears more
suitable for our variable-free analysis.

\subsubsection{Comparison of (\ref{veca1}) and (\ref{veca2})}

\label{secMODdescrip1}

Using our illustrative  gene regulatory
network example (\ref{prodP1P2}) -- (\ref{onoffO1O2})
we now study the dependence of
the eigenvectors of the matrix $\boldK$ on the weighting vector
$[a_1,a_2, \dots, a_N]$.
We run the long-time Gillespie based stochastic
simulation of (\ref{prodP1P2}) -- (\ref{onoffO1O2})
to obtain a representative set of $M$ state vectors
using
the following dimensionless stochastic rate constants
$
\gamma_1 = \gamma_2 = 1.14,
$ $
\varepsilon_1 = \varepsilon_2 = 0,
$ $
\delta_1 = \delta_2 =7.5 \times 10^{-4},
k_1 = k_2 = 10^{-3},
$ $
k_{-1} = k_{-2} = 10,
$ $
k_{o1} = k_{o2} = 0.4,
$ $
k_{-o1} = k_{-o2} = 10.
$
After removing initial transients, we started recording
the values of the state vector (\ref{statevectormutrep})
every $2 \times 10^8$ SSA time steps.
We made $2000$ recordings
to obtain a data file with $M=2000$ state vectors.
Next, we use these state vectors $\boldX^{(i)}$
to compute the $M \times M$ matrix $\boldK$ and its eigenvectors.
We use formula (\ref{veca1}) to compute $\boldW$ and $\boldD$ by
(\ref{weightmatrixnoeucl}), (\ref{weightmatrix}) and (\ref{diagmatrix}).
Then we use implicitly restarted Arnoldi methods (ARPACK package
\cite{Lehoucq:1998:AUG})
to find the eigenvectors corresponding
to the highest eigenvalues of the
symmetric matrix $\boldS$ given by (\ref{symmatrix}).
Finally, we compute the eigenvectors of
$\boldK = \boldD^{-1} \boldW$ by (\ref{eigenK}).

The formula (\ref{veca1}) has a single parameter $\omega$ which
is free for us to specify.
It is easy to check numerically that the larger the ``local
neighborhood'' size selected (that is, the smaller the $\omega$ value) the
denser the connections between datapoints in the graph.
Table \ref{tab1} shows the highest eigenvalues for different
values of $\omega$.
\begin{table}
\centerline{
\begin{tabular}{|c|c|c|c|c|} \hline
$\omega$ & $\lambda_0$  & $\lambda_1$ & $\lambda_2$ & $\lambda_3$ \\
\hline
0.02 & 1.00000 & 0.99986 & 0.94506  &  0.91360 \\
\hline
0.01 & 1.00000 & 0.99920 & 0.77757  &  0.71122 \\
\hline
0.005 & 1.00000 & 0.99279 & 0.44352  &  0.35515 \\
\hline
0.002 & 1.00000 & 0.76262 & 0.10715  &  3.3 $\times 10^{-2}$ \\
\hline
0.001 & 1.00000 &  0.28346 & 1.2 $\times 10^{-2}$  &  1.1 $\times 10^{-3}$ \\
\hline
0.0005 & 1.00000 &  7.5 $\times 10^{-2}$ & 1.0 $\times 10^{-3}$  &  1.5 $\times 10^{-4}$ \\
\hline
\end{tabular}}
\caption{{\it Top eigenvalues of matrix $\boldK$ computed using
    $(\ref{veca1})$ for $\alpha=0$ in $(\ref{weightmatrix})$.} }
\label{tab1}
\end{table}
We already know from \cite{Erban:2006:GRN} that the system is
effectively one-dimensional.
A good observable for
the system is known to be $Q=P_1-P_2$, i.e. the difference between the
first two coordinates of the state vector.
However,
the protein concentrations
$P_1$ or $P_2$ were also found to give good equation-free results.

\begin{figure}
\centerline{{\Large $\omega = 0.002$}
\qquad \qquad \qquad \qquad \qquad \qquad {\Large $\omega = 0.0005$}}
\centerline{
\psfig{file=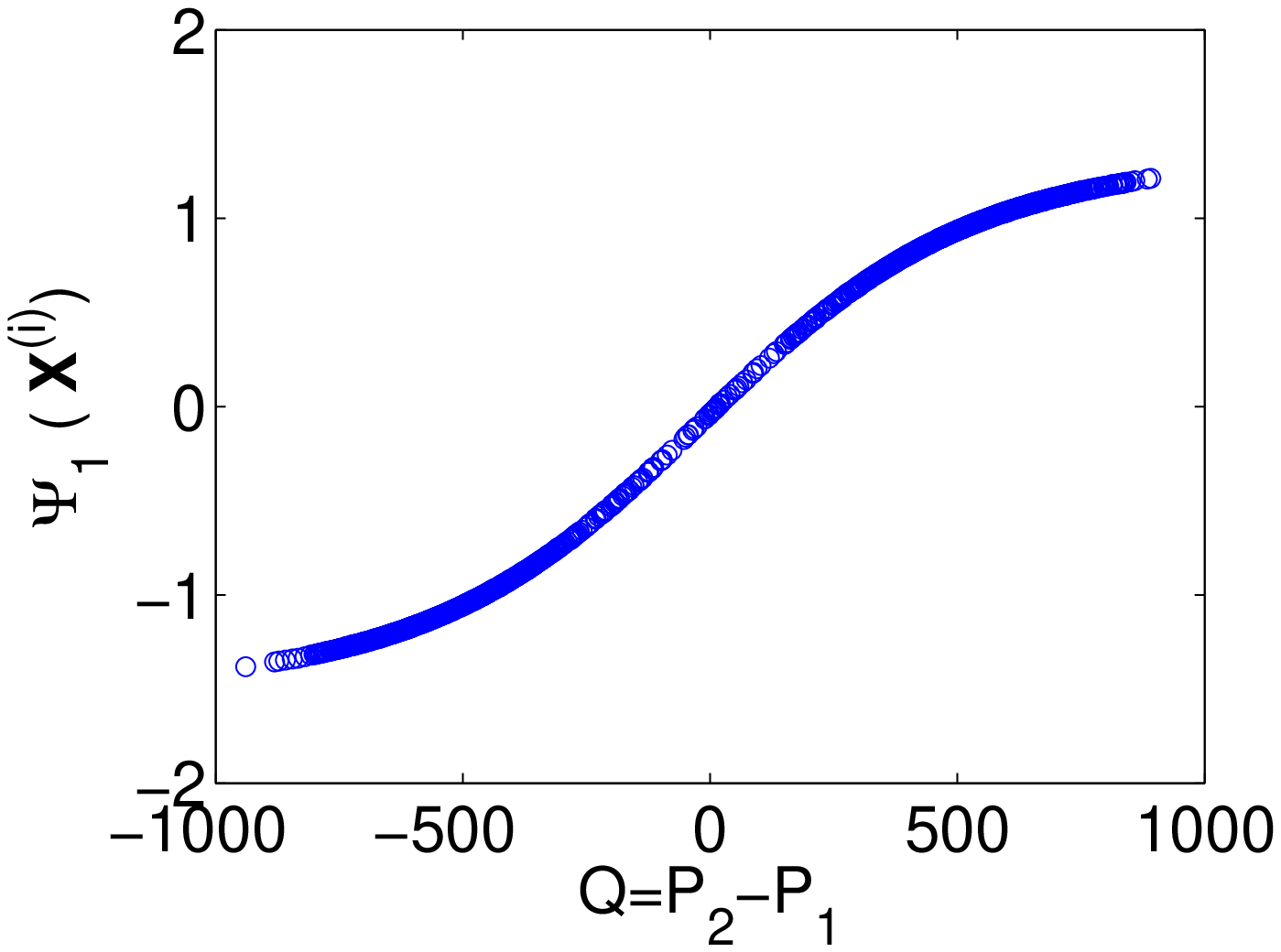,height=2.2in}
\quad
\psfig{file=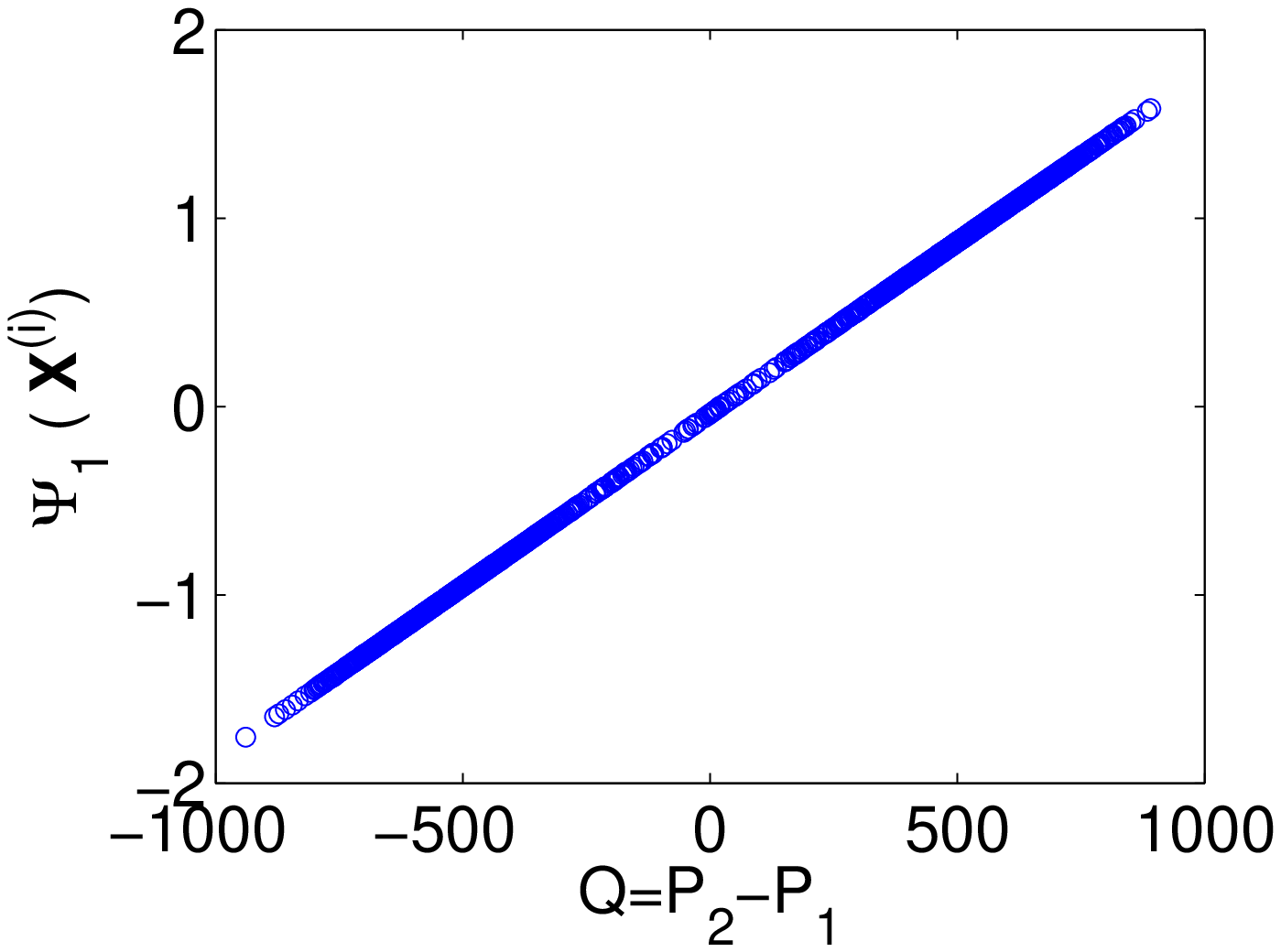,height=2.2in}
}
\smallskip
\centerline{
\psfig{file=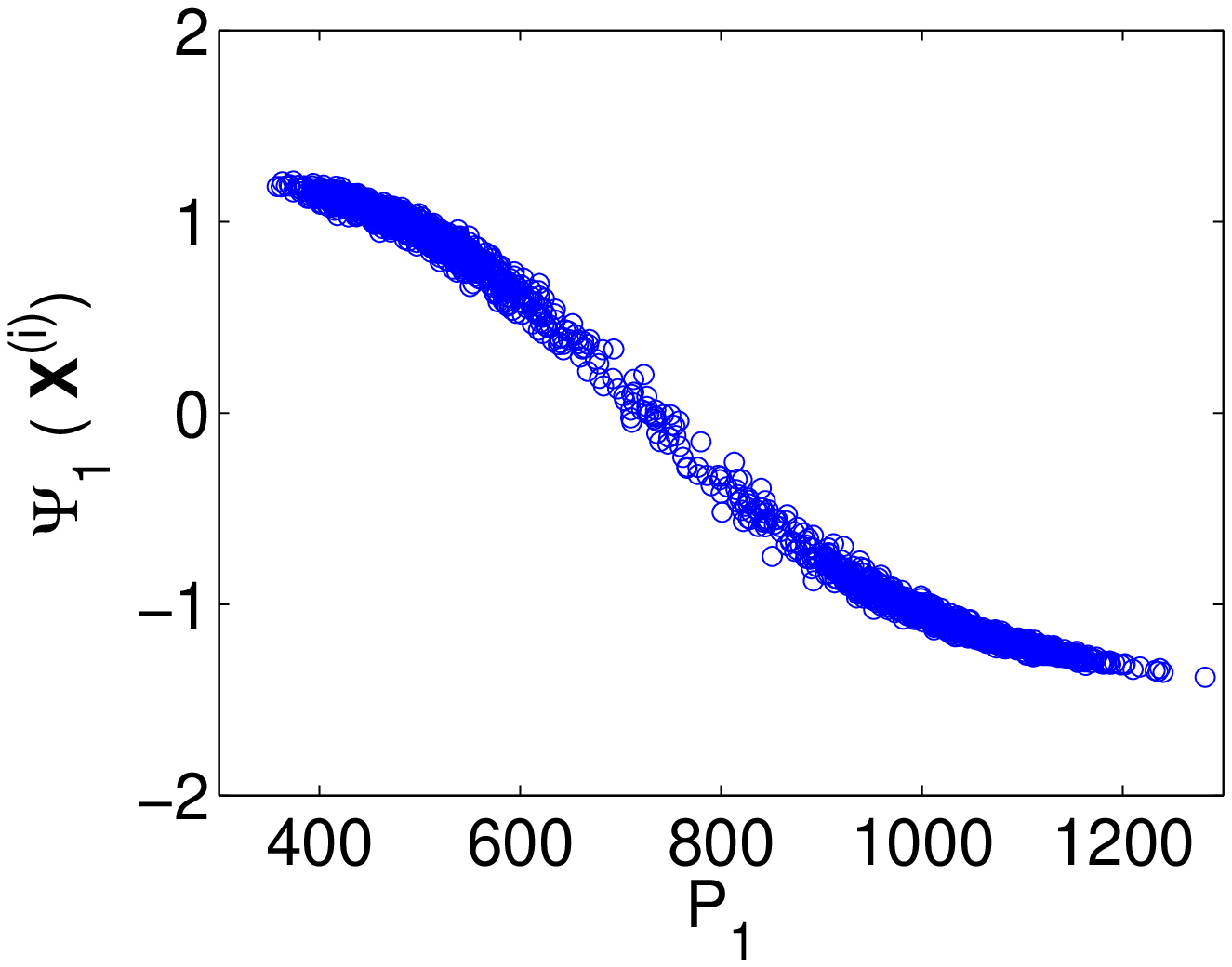,height=2.2in}
\quad
\psfig{file=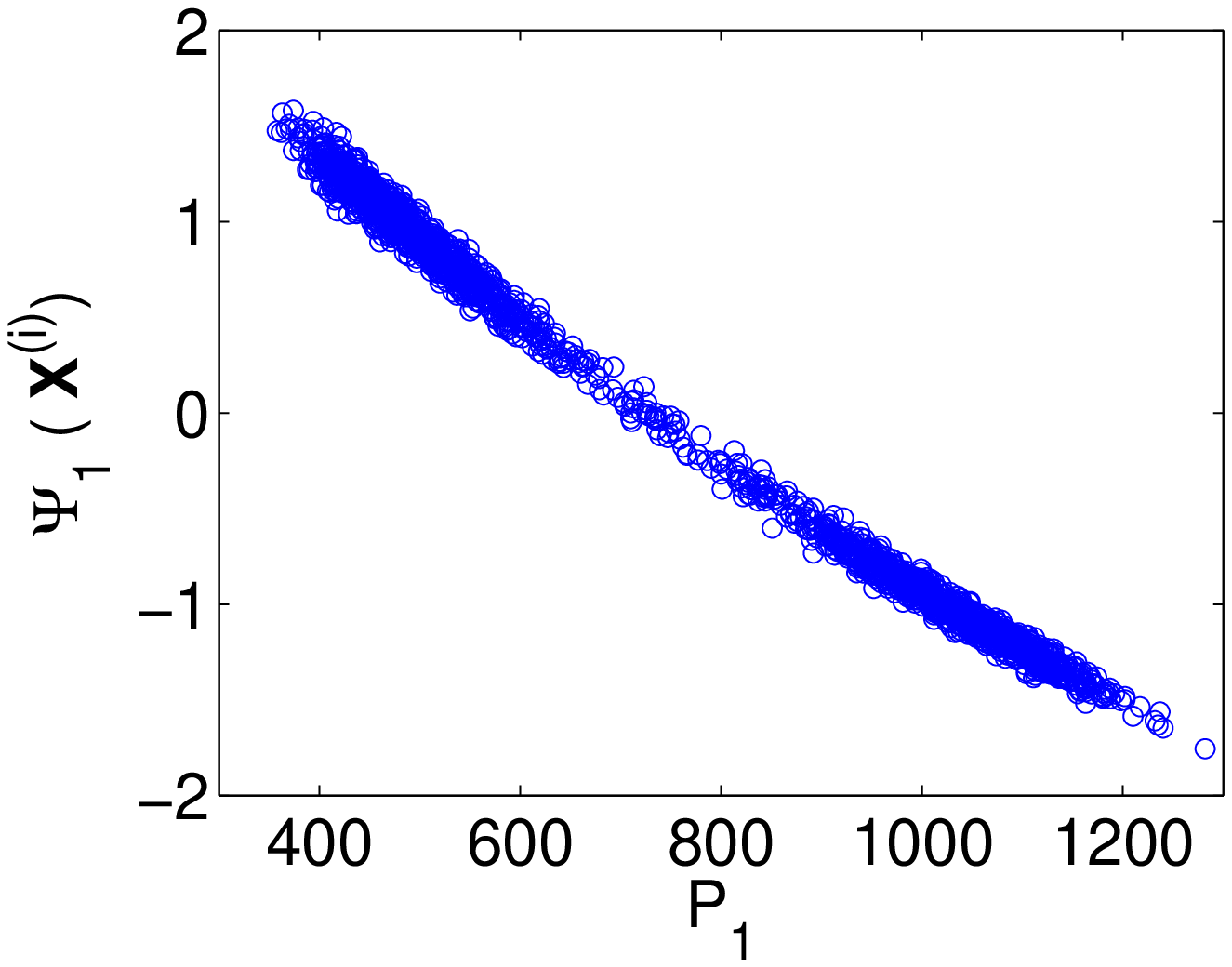,height=2.2in}
}
\smallskip
\centerline{
\psfig{file=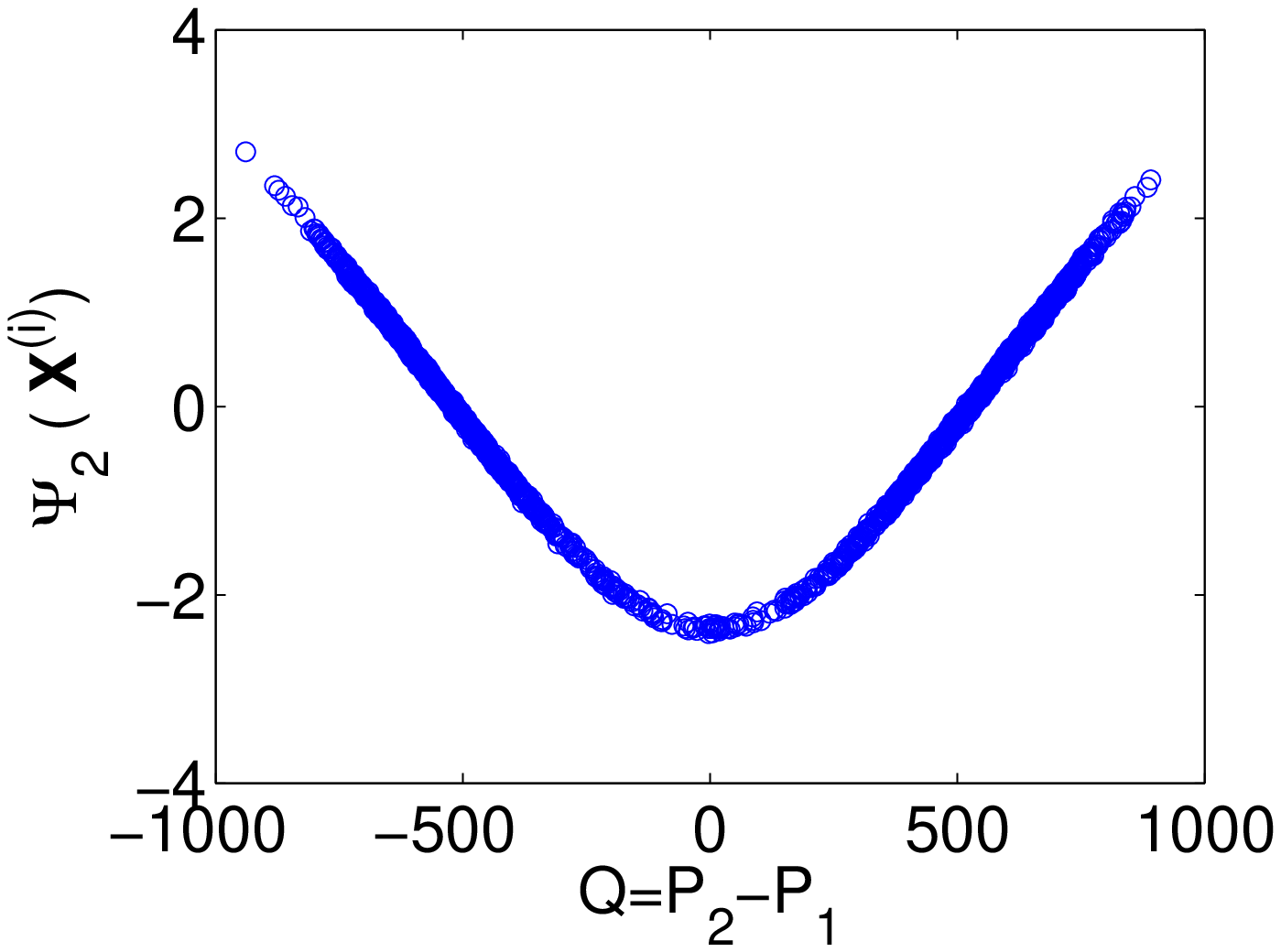,height=2.2in}
\quad
\psfig{file=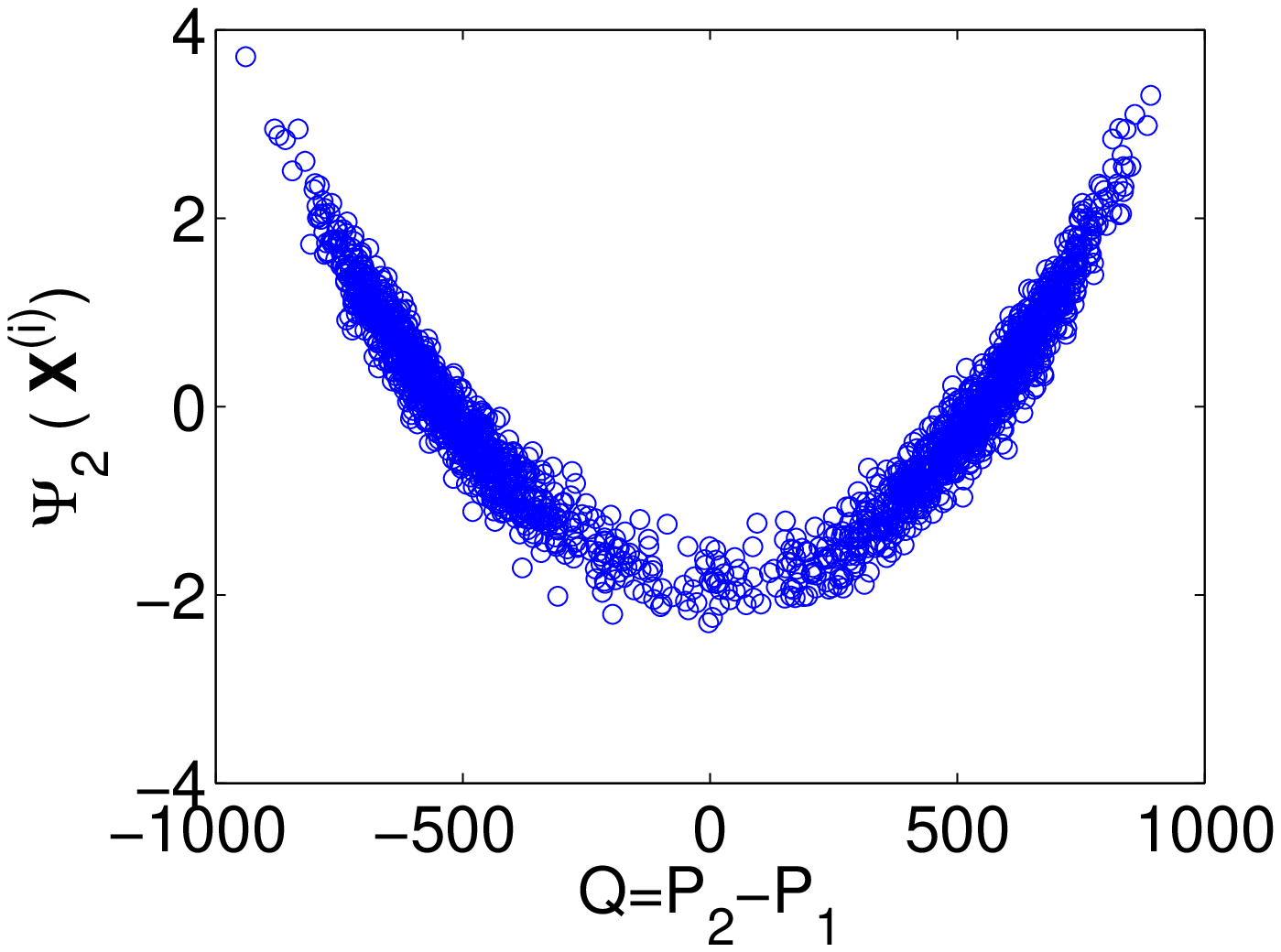,height=2.2in}
}
\caption{{\it Variable-free results using formula
$(\ref{veca1})$ and $\omega=0.002$ (left panels)
or $\omega=0.0005$ (right panels).
We plot $\Psi_1 (\boldX^{(i)})$
which corresponds to eigenvalue $\lambda_1$
as function of $Q=P_2-P_1$ (top panels)
and as function of $P_1$ (center panels).
We also plot $\Psi_2 (\boldX^{(i)})$
which corresponds to eigenvalue $\lambda_2$
as function of $Q=P_2-P_1$ (bottom panels).
}}
\label{figsymmeigen2}
\end{figure}

We plot the ``empirical" good observable of each data point $i$
(its $P_1$ component, i.e. $X_1^{(i)}$, or the difference of its $P_1$ and
$P_2$ components, i.e. $Q = X_2^{(i)}-X_1^{(i)})$ versus the one-dimensional
representation $\Psi_1 (\boldX^{(i)})$ (see
(\ref{lowdimenrepresentation}))
 of the point.
The results are given
in Figure \ref{figsymmeigen2} for two different values of $\omega$.
The fact that the empirical coordinate $Q$ appears to effectively
be {\em one-to-one} with the ``automated" coordinate
$\Psi_1 (\boldX^{(i)})$ for all
points in the data set confirms that $Q$ is indeed a good
coordinate for data representation (the figure clearly shows $Q$
as the graph of a function above $\Psi_1 (\boldX^{(i)})$, i.e. that the relation
between $Q$ and $\Psi_1 (\boldX^{(i)})$ is one-to-one).
The $P_1$ vs. $\Psi_1 (\boldX^{(i)})$ graph confirms that
$P_1$ is also a good observable; it also is approximately one-to-one with $\Psi_1 (\boldX^{(i)})$,
yet the slightly ``fat curve" suggests that $Q$ is
a ``better" observable.

The dependence of the variable-free results on the value chosen for $\omega$
may be rationalized through equation (\ref{weightmatrixtilda0}).
As discussed in Section \ref{secVFnumer},
our parameter $\omega$ is analogous to an inverse ``cutoff length"
in the computation of the diffusion map kernel; if it is too large,
then the graph becomes disconnected.
%
Clearly, it is a model parameter that has to be optimized depending on the
problem; our results for $\omega=0.0005$ show a pure
linear relation between the ``empirical" $Q$ and the ``automated"
$\Psi_1 (\boldX^{(i)})$ observables.
Increasing $\omega$ by a factor of 2 corresponds to raising the elements of the
matrix $\widetilde{\boldW}$ to the fourth power.
This change in weight factor (followed by the normalization of (\ref{diagmatrix}))
leads to a different clustering of the data points.
Large $\omega$ implies that Euclidean distances are meaningful when small;
this results in a ``more clustered" data set, where nearby data points
(e.g. points within one potential well) appear (in diffusion map
coordinates) relatively closer, while
points far away (e.g. points in different potential wells) appear
(in diffusion map coordinates) relatively more distant.
Indeed, in the case of continuous variables, in the limit
of large $\omega$ the eigenvectors of the diffusion map
converge to the eigenfunctions of a corresponding Fokker-Planck diffusion
operator. 
In the case of two deep potential wells,
this eigenfunction is approximately constant in the two wells with
a sharp transition between them.
This might explain the slightly flat regions at the two edges of
the apparent curve in the middle panel of Figure \ref{figsymmeigen2}
for $\omega=0.002$; points within the same potential
well may differ in $Q$, yet appear more nearby in the ``automated"
observable.
We also include a plot of the relation between $Q$ and the component of the
data in the second eigenvector $\Psi_2 (\boldX^{(i)})$ for comparison.

\begin{figure}
\centerline{{\Large $\widetilde\omega = 2$}
\qquad \qquad \qquad \qquad \qquad \qquad \qquad {\Large $\widetilde\omega = 0.1$}}
\centerline{
\psfig{file=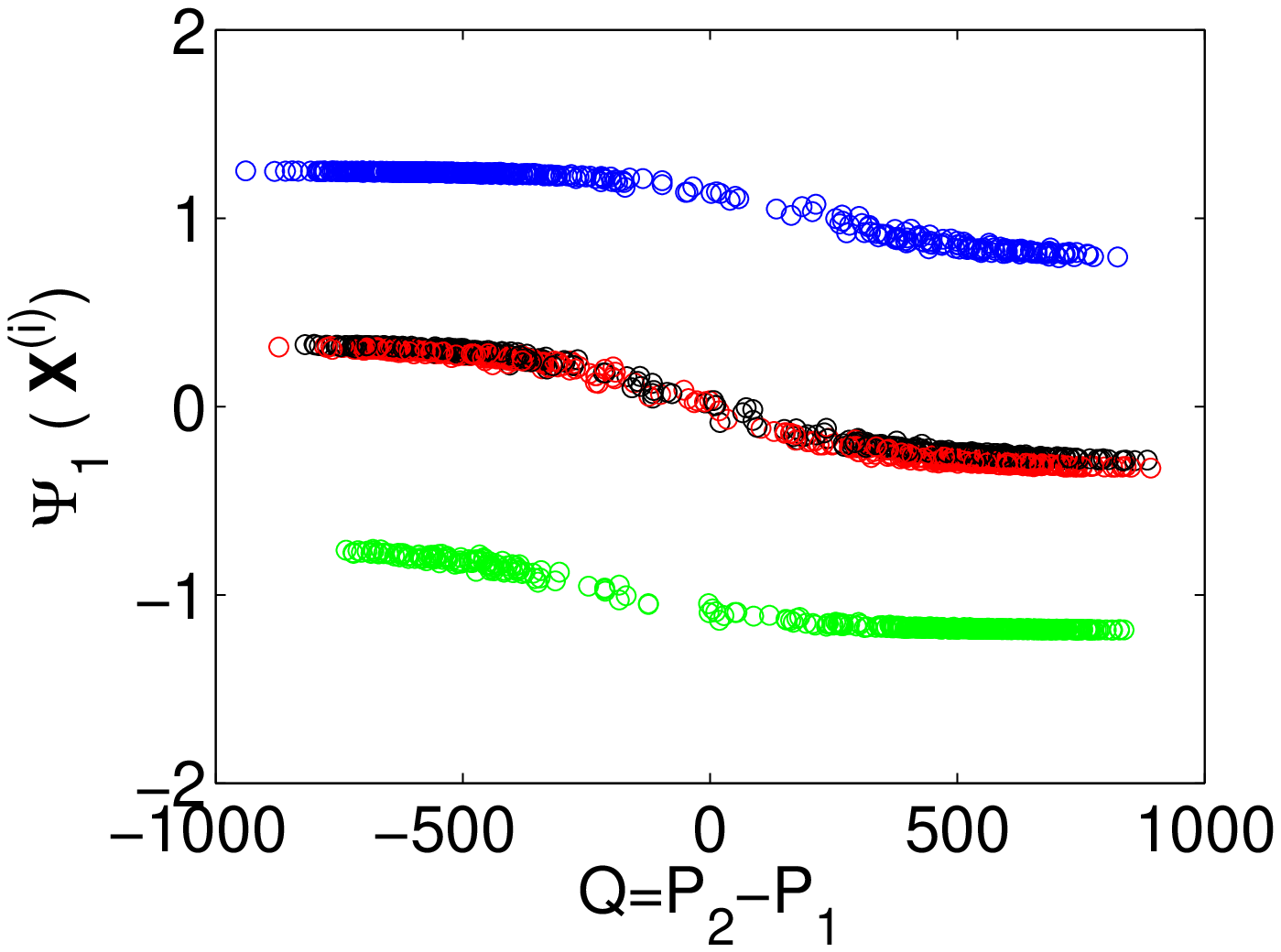,height=2.2in}
\psfig{file=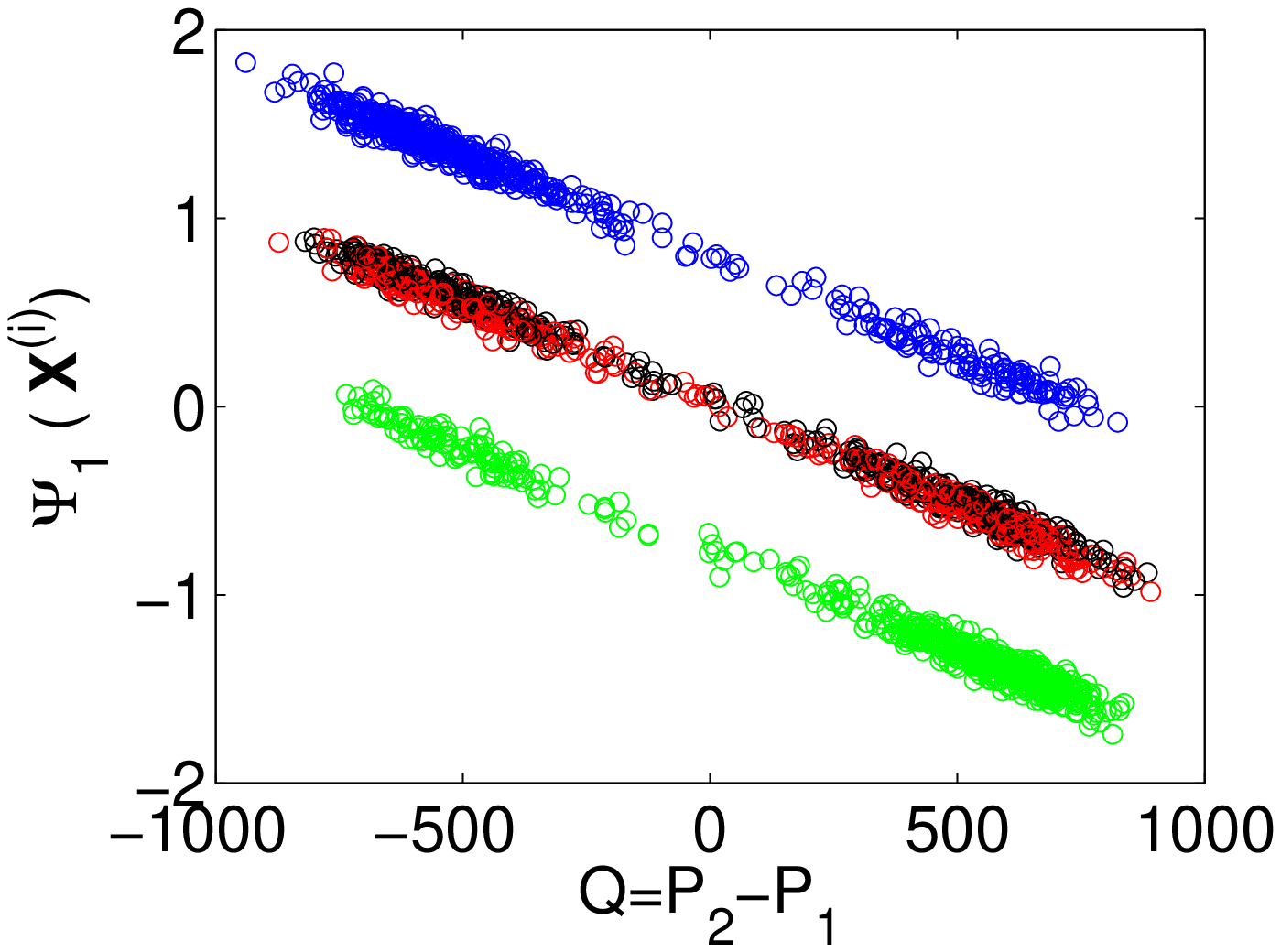,height=2.2in}
}
\caption{{\it Variable-free results using formula
$(\ref{veca2})$ and $\widetilde\omega=2$ (left panel)
or $\widetilde\omega=0.1$ (right panel); datapoints colored according to gene
states: black=$[0,0]$, green=$[0,1]$, blue=$[1,0]$, and red=$[1,1]$.
We plot $\Psi_1 (\boldX^{(i)})$
which corresponds to eigenvalue $\lambda_1$
as function  of $Q=P_2-P_1$.
}}
\label{figsymmeigenscaled2}
\end{figure}

Next we show that the weight vector computed using the formula (\ref{veca2})
(based on the magnitude of {\em relative} state variable changes)
is unsuitable for our variable-free analysis.
We use  the same set of $M=2000$ state vectors
$\boldX^{(i)}$  to compute the $M \times M$ matrix $\boldK$ and its
eigenvectors, using formula (\ref{veca2}) to compute $\boldW$ and
$\boldD$ by (\ref{weightmatrixnoeucl}), (\ref{weightmatrix}) and 
(\ref{diagmatrix}).
A single parameter $\widetilde\omega$ still remains to be specified in formula (\ref{veca2}).
%
%
%
We now again compare the ``empirical" and ``automated" observables
of all data points ($Q=P_1-P_2$ as a function of $\Psi_1 (\boldX^{(i)})$, the
one-dimensional representation based on the first nontrivial
eigenvector of the matrix $\boldK$).
The results are given in Figure
\ref{figsymmeigenscaled2} for two different values of $\widetilde\omega$.
We see that the data split into four curves.
Each curve corresponds to a distinct combination
of gene operator states (actually, two of the
curves effectively coincide).
There are exactly four possibilities
of gene states taken from the set
$$
[O_1,O_2] \in \Big\{ [0,0], \; [0,1], \; [1,0] \; [1,1] \Big\}.
$$
If we use formula (\ref{veca1}), then the contribution of
the distance
between gene operator states to the data Euclidean distance
is negligible compared to the fluctuations
of the protein numbers.
Local distances computed using the scaling in formula (\ref{veca2})
are clearly {\em not} representative of the similarity of
nearby (in this metric) points for the system dynamics: there is
no one-to-one correspondence between the
empirically known ``good observable" $Q$ and the ``automated"
$\Psi_1 (\boldX^{(i)})$ .
Indeed, for the parameter values of our simulation, transitions
between the 0 and 1 states of the operators are {\em very} fast
(``easy"); on the other hand the Euclidean distance of two
data points that differ only in these states is large when
computed through the formula (\ref{veca2}).

An alternative approach to computing the effective rate in (\ref{prodP1P2})
can be obtained assuming that the reaction (\ref{prodP1P1P2P2})
is fast and that we have a lot of protein molecules in the system.
Then the quasi-steady state assumption gives the formula
$\overline{P_1P_1} = 2 k_1/k_{-1} P_1^2$.
Hence, we can write the number of dimers as a simple function
of the number of monomer proteins.
On the other hand, using the same approximation
in equation (\ref{onoffO1O2}), we obtain
\begin{equation}
O_1 = \frac{k_{-o1}}{k_{o1} \overline{P_2P_2} + k_{-o1}}.
\label{ststex}
\end{equation}
Equation (\ref{ststex}) gives
$O_1$ as a real number in the interval $[0,1]$.
This number
is a good approximation for computing the effective rate
in (\ref{prodP1P2}).
However, it is not a value {\em of the Boolean variable} $O_1$
-- it is only a probability that the gene ``is on" at the
given time.

If, on the other hand, the ``on-off" operator transitions were slow,
then Figure \ref{figsymmeigenscaled2} would be quite informative: it would suggest that we should
{\it augment} our observables with the Boolean variables $O_1$, $O_2$,
since these are ``slow".
Because of the Boolean nature of the gene operator variables, it is not
possible to know {\em a priori} how often these transitions occur, and,
consequently, how to scale the quantized Boolean state distance so that it
``meaningfully" participates in the Euclidean distance used for
diffusion map analysis.
As our diffusion map computations stand, we do not take into account the
{\it temporal} proximity of points -- when they have been obtained from
the same transient.
If such information is taken into account, it is
conceivable that temporal proximity would provide guidance in choosing
the components of weight vectors (especially for Boolean variables which change in
a quantized manner) so that ``local" Euclidean distances are indeed representative
of the dynamical proximity between data points.

\section{Variable-free computations}

\label{secVarFree}
We now couple the above automated detection of observables
with the equation-free computations in \cite{Erban:2006:GRN} in what
we will refer to as ``variable-free, equation-free"  methods.
The results in this section are for the model parameter values given in
Section \ref{secMODdescrip1} using the weight vector defined by
(\ref{veca1}) with $\omega=0.0005$ and kernel parameter $\alpha=0$ (the standard,
normalized graph Laplacian) in (\ref{weightmatrix}).

The data plot in terms of the observable $Q$ and the component in the
eigenvector $\Psi_1 (\boldX^{(i)})$ in Figure \ref{figsymmeigen2}
suggested that a single diffusion map coordinate, denoted
$Q_{dmap}\equiv \Psi_1 (\boldX^{(i)})$,
is sufficient to characterize the system dynamics.
The diffusion map coordinate is found by performing the
eigencomputations described in Section \ref{secVFtheory} using the full state
vector ($N$=6) at each of the $M=2000$ recorded
SSA datapoints (every $2\times 10^8$ SSA time steps) as input to our numerical routines.
%

\begin{figure}
\centerline{
\psfig{file=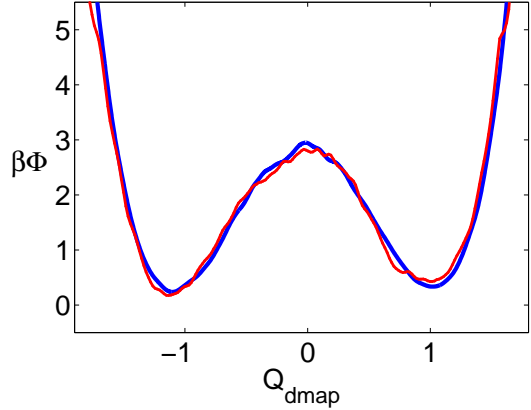,height=2.3in}
}
\caption{{\it Effective free energy $\beta\Phi$ as a function of
    $Q_{dmap}$ from binning of all datapoints using 
    an SSA database of $2^{37}$ time steps (blue lines) and computed
    from  numerical integration of equation
    $(\ref{potentialPhi})$ using a $2^{34}$ point subsampling (keeping
    $1$ out of every $8$ points) of this database (red lines). Numerical integration performed using a more
    severe subsampling of the database with $2^{31}$ points produces an effective free
    energy profile with an unacceptable level of noise.}}
\label{figsEFdmap}
\end{figure}

In our previous paper \cite{Erban:2006:GRN} we described an approach
to compute an effective free energy potential in terms of the
observable $Q=P_2-P_1$.
Variable-free computation of the effective free energy
is now feasible using a similar approach modified to analyze
simulation data in terms of the coordinate $Q_{dmap}$.
%
Figure \ref{figsEFdmap} plots the effective potential $\beta\Phi$ in
terms of the automated reduction coordinate $Q_{dmap}$. 
To evaluate the
effective drift ($V$) and diffusion ($D$) coefficients required in the construction of the effective free energy (equation (\ref{potentialPhi})) we choose a value of $Q_{dmap}$, locate instances when it appears in the simulation database, record its subsequent evolution within a fixed time interval, and then average over these instances to estimate the rate of change in the
mean and the variance. 
This procedure is repeated for a grid of
$Q_{dmap}$ values enabling numerical evaluation of the integral in
equation (\ref{potentialPhi}). 
The result of this analysis is compared in
Figure \ref{figsEFdmap} with the potential obtained by directly
constructing the probability distribution $f(Q_{dmap})$
from the time series and employing the relationship
$\beta\Phi(Q_{dmap})\sim-\log\left[f(Q_{dmap})\right]$.

Section \ref{secLIFTING} describes a {\it lifting} procedure that allows short bursts
of simulation, instead of long time simulation, to be used 
in variable-free estimation of effective drift and diffusion coefficients.
The central idea of ``variable-free equation-free" methods is to perform 
equation-free analysis
{\em in terms of diffusion map variables}, based on short bursts of SSA simulation
{\em in the original variables}.
This strategy requires an efficient means of converting
between the physical variables of the system and those of its
diffusion map (a {\it restriction} step) and vice versa: {\it lifting}
from the diffusion map back to physical variables.
For small sample sizes,
eigendecomposition of the symmetric kernel $\boldS$ (defined in (\ref{symmatrix})) yields the
diffusion map variables {\it for each data point}; yet, as the number of sample datapoints
increases, the associated computational costs become prohibitive.
The Nystr\"{o}m formula \cite{Baker:1977:Tnt,Bengio:2004:LE} for eigenspace interpolation is a viable
alternative to repeated matrix eigendecompositions for computing
diffusion map coordinates of new  datapoints generated during the
course of a simulation.
Eigenvectors and eigenvalues of the  kernel
$\boldS$  are related by $\boldS \boldU_j = \lambda_j \boldU_j$,
or equivalently
\begin{equation}
U_j(\boldX^{(i)})=\frac{1}{\lambda_j}\sum_{k=1}^{M} S_{ik}U_j(\boldX^{(k)})
\label{eigexpans}
\end{equation}
where $U_j(\boldX^{(i)})$ denotes the component of the $j^{th}$
eigenvector associated with state vector $\boldX^{(i)}$.
Eigenvector components associated with a {\em new} state vector $\boldX^{new}$
cannot be computed directly from (\ref{eigexpans}) because entries of
the matrix $\boldS$ are defined only between pairs of datapoints in the original dataset.
Defining the $M \times 1$ vector $\widehat{\boldW}^{new}$ of exponentials
of the negative squares of the distances between the new point and
database points by

\begin{equation}
\widehat{W}_{i}^{new}=\exp\big[-\norm{\boldX^{new} - \boldX^{(i)}}_{\bf a}^2\big],
\label{weightvec}
\end{equation}
and the $M \times 1$ vector ${\boldW}^{new}$ by
\begin{equation}
W_{i}^{new}
=
\left( \sum_{k=1}^M \widetilde{W}_{ik} \right)^{-\alpha}
\left( \sum_{k=1}^M \widehat{W}_{k}^{new} \right)^{-\alpha}
\widehat{W}_{i}^{new},
\label{weightvec2}
\end{equation}
allows the generalized kernel vector $\boldS^{new}$ to be defined
as follows:

\begin{equation}
S_{i}^{new}=
\left(\sum_{k=1}^{M} W_{ik}\right)^{-1/2}\left(\sum_{k=1}^{M}{W}_{k}^{new}\right)^{-1/2}{W}_{i}^{new}.
\label{kernelvec}
\end{equation}
The entries in $\boldS^{new}$ quantify the pairwise similarities between the
new point $\boldX^{new}$ and database points consistent with the definition of $\boldS$
in (\ref{symmatrix}) \cite{Bengio:2004:LE}.

\subsection{{\it Restriction} from physical to diffusion map variables}

\label{secRESTRICTING}

The Nystr\"{o}m
formula \cite{Baker:1977:Tnt} is used to find the eigenvector component $U_j(\boldX^{new})$ associated with a
new state vector $\boldX^{new}$

\begin{equation}
U_j(\boldX^{new})=\frac{1}{\lambda_j}\sum_{i=1}^{M}\widehat{S}_{i}^{new} U_j(\boldX^{(i)})
\label{nystrom}
\end{equation}
allowing the eigenvectors of the matrix $\boldK$ (and thereby the
diffusion map coordinates) associated with $\boldX^{new}$ to be computed using
(\ref{eigenK}).
A full eigendecomposition is typically performed first for a representative subset of
the (large) number of SSA datapoints and the Nystr\"{o}m formula is then used to
perform the restriction operation in (\ref{nystrom}) which amounts to
interpolation in the diffusion map space.

\subsection{{\it Lifting} from diffusion map to physical variables}

\label{secLIFTING}

\begin{figure}
\centerline{
\psfig{file=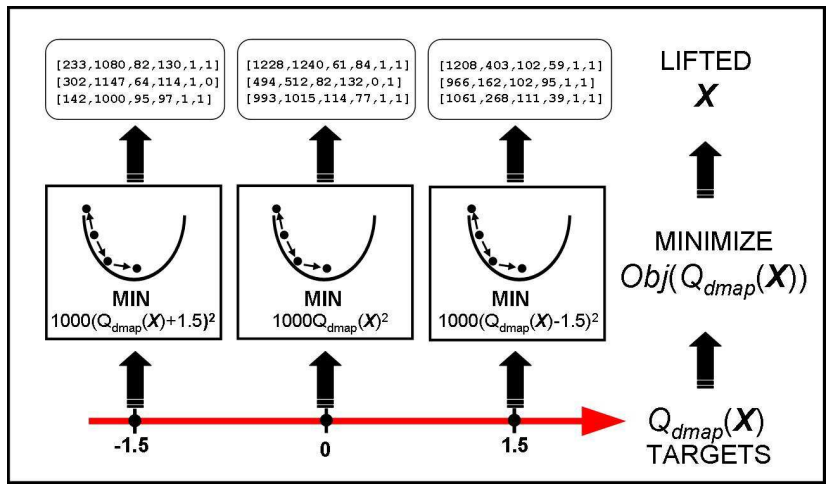,height=2.5in}
}
\caption{{\it A schematic of the procedure for lifting from diffusion
    map coordinate $Q_{dmap}(\boldX)$ to 6-dimensional state vector
    $\boldX$ via minimization of quadratic constraint potential
    $Obj(Q_{dmap}(\boldX))$.
    Target values of diffusion map coordinate
    are shown at the base of the figure, with  the potential function
    to be minimized in each case
    indicated above these targets.
    For each diffusion map coordinate
    value shown, $3$ consistent state vectors (generated by lifting) are indicated at
    top of figure.}}
\label{figlifting}
\end{figure}

The process of {\it lifting} (shown
schematically in Figure \ref{figlifting}) consists of preparing a
detailed state
vector with prescribed diffusion map coordinates $Q_{dmap}^{targ}$.
The main step in our
lifting process is the minimization of a quadratic objective function
defined as follows

\begin{equation}
Obj(Q_{dmap}(\boldX))=\lambda_{obj}(Q_{dmap}(\boldX)-Q_{dmap}^{targ})^2
\label{objfunction}
\end{equation}
where $\lambda_{obj}$ is a weighting parameter that controls the
shape of the objective away from its minimum at
$Q_{dmap}(\boldX^*)=Q_{dmap}^{targ}$.
The implicit dependence of
$Q_{dmap}$ on $\boldX$ makes this optimization problem nontrivial.

We use here, for simplicity, the method of Simulated
Annealing (SA) \cite{Kirkpatrick:1983:Sci,Press:1992:NR} to solve the
optimization problem, and identify a value of the
state vector $\boldX^*$ with the target diffusion map coordinates
$Q_{dmap}^{targ}$.
The SA routine \cite{Press:1992:NR} employs a ``thermalized'' downhill simplex method
as the generator of changes in configuration.
The simplex, consisting
of $N+1$ vertices, each corresponding to a trial state vector, tumbles over
the objective landscape defined by (\ref{objfunction}) sampling new
state vectors as it does so.
The control parameter of the method is the
``annealing temperature" which controls the rate of simplex
motion.
At high temperatures the method behaves like a global
optimizer, accepting many proposed configurations (even those that
take the simplex uphill i.e. in the direction of increasing objective function
value).
At low temperatures a local search is executed and only
downhill simplex moves are accepted.

The starting simplex configuration for this $N$-parameter minimization may be
selected at random or (more reasonably) by taking those state vectors in the existing database
with diffusion map coordinates closest to the target
$Q_{dmap}^{targ}$.
It is important to note that
the SA optimization scheme {\it requires} the Nystr\"{o}m
formula  at each iteration to compute $Q_{dmap}(\boldX^{trial})$ for trial state vectors,
and thus evaluate the objective function value, which determines whether
the configuration will be accepted or not.
Once the objective  has been evaluated at each of the starting vertices,
the following steps are repeated until a minimum is located:

\leftskip 1cm
\bigskip

\noindent
({\it a}) Move the simplex to generate a new state vector $\boldX^{trial}$;

\noindent
({\it b}) Evaluate the objective function value at the new state
  vector  $Obj(Q_{dmap}(\boldX^{trial}))$;

\noindent
({\it c}) Decrement the annealing temperature.

\leftskip 0cm
\bigskip

\noindent
The downhill simplex method prescribes the motion in
step ({\it a}) making a selection from a set of moves according to the
local objective ``terrain'' (set of objective values at the vertices
encountered).
Step ({\it b}) requires an evaluation using the
Nystr\"{o}m formula. We note here that this lifting strategy
 prepares state vectors with desired diffusion map coordinates
using search algorithm ``dynamics''. The suitability of this
approach relative to alternatives that employ physical dynamics
(e.g. using constrained
evolution of the stochastic simulator in the spirit of the SHAKE
algorithm in molecular dynamics \cite{Ryckaert:1977:NIC}) is a
relevant and interesting question that merits further investigation.

\subsection{Illustrative Numerical Results}

Equipped with restriction and lifting operators between physical and
``automated'' variables, we can now perform all the equation-free tasks
of \cite{Erban:2006:GRN} in the diffusion map
coordinate $Q_{dmap}$ i.e. in variable-free mode.

A procedure for variable-free computational estimation of $V(q)$ and
$D(q)$ in (\ref{potentialPhi}) is as follows:

\leftskip 1cm
\bigskip

\noindent
{\bf [A]} At the value $Q_{dmap}=q_{dmap}$ lift to a consistent state vector
using the approach described in Section \ref{secLIFTING}.

\noindent
{\bf [B]} Use the state vector computed in step [A] as an initial
condition for a short simulation burst and run multiple realizations
for time $\Delta t$.
Restrict the results of these simulations (Section
\ref{secRESTRICTING}) and use the definitions (\ref{avgvel}) and
(\ref{effdiff}) (with $Q_{dmap}(t)$ instead of $Q(t)$) to estimate
the effective drift $V(q_{dmap})$ and the effective diffusion coefficient
$D(q_{dmap})$.

\noindent
{\bf [C]} Repeat steps [A] and [B] for sufficiently many values of $Q_{dmap}$ and then compute
$\Phi(q)$ using formula (\ref{potentialPhi}) and numerical quadrature.

\leftskip 0cm
\bigskip

\noindent
\begin{figure}
\centerline{
\psfig{file=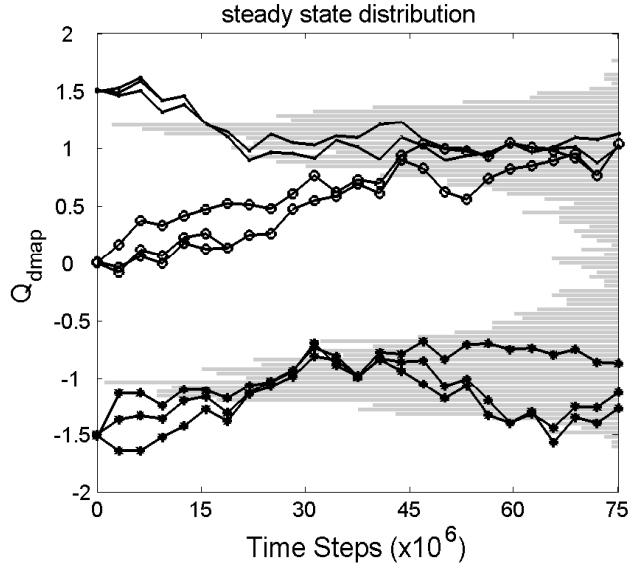,height=3.in}
}
\caption{{\it Drift in the diffusion map coordinates. The shaded
    horizontal boxes
    indicate the steady state probability distribution for $M=2000$. Points from SSA trajectories are shown at intervals of $3\times 10^{6}$. Initial configurations for these
    runs are those shown in Figure $\ref{figlifting}$ prepared
    by lifting from $Q_{dmap}$ values of $(-1.5,0,1.5).$ Trajectories
    drift towards the most populated regions of the distribution.}}
\label{figdmapdrift}
\end{figure}

\noindent We performed lifting for 3 values of the automated reduction coordinate
($Q_{dmap}=-1.5,0,1.5$), generating several replicas in each case.
From Figure \ref{figdmapdrift} it is apparent that the
selected values of $Q_{dmap}$ are located near the ``rims" of the wells of
two local
minima on the effective free energy landscape for this system.
The state vectors generated
by lifting are shown at the top of Figure \ref{figlifting}.
Figure
\ref{figdmapdrift} plots the SSA simulation evolution,
initialized at these state vectors, in the observable $Q_{dmap}$.
Also shown in Figure \ref{figdmapdrift} is the steady state
distribution in terms of $Q_{dmap}$ obtained from long SSA runs.
Estimates for drift ($V$) and diffusion ($D$) coefficients at
$Q_{dmap}$ values of -1.5 and 0 produced by sampling the
simulation database and using the lifting procedure described in this
paper are compared in Table \ref{tabv+D}. 
It should be possible to reach a better agreement between the
coefficient estimates based on the long simulation database and those
obtained by a lifting procedure if we evolve the actual model dynamics
with a constraint on the prescribed $Q_{dmap}$ value - possibly through
a parabolic constraint potential of the type used in umbrella sampling
(see also the ``run and reset" procedure described in 
\cite{Erban:2006:EFC,Erban:2006:GRN}). 
%
%
The effective free energy predicted by
analyzing the full simulation database in terms of $Q_{dmap}$ can be
found in
Figure \ref{figsEFdmap}. 
\begin{table}
\centerline{
\begin{tabular}{|c|c|c|c|c|} \hline
  & \multicolumn{2}{|c|}{Database}  & \multicolumn{2}{|c|}{Lifting} \\
\hline
 $(Q_{dmap})_0$ & $V$ & $D$ & $V$ & $D$ \\
\hline
-1.5 & 3.3 $\times 10^{-5}$ & 4.7 $\times 10^{-6}$ & 2.1 $\times
10^{-5}$ & 3.2 $\times 10^{-6}$  \\
\hline
0. & 5.3 $\times 10^{-6}$ & 4.0 $\times 10^{-6}$ & 7.9 $\times
10^{-8}$ & 4.1 $\times 10^{-6}$  \\
\hline
\end{tabular}}
\caption{{\it Estimates for drift ($V$) and diffusion ($D$)
    coefficients (in $s^{-1}$) at $Q_{dmap}$ values of $-1.5$ and $0$ 
    using initial conditions drawn from the simulation database and 
    prepared by lifting.} }
\label{tabv+D}
\end{table}

\section{Summary and Conclusions}

The knowledge of good observables is vital in our ability to create
effective reduced models of complex systems, and thus to analyze and even design their
behavior at a macroscopic/engineering level more efficiently.
In this paper we illustrated a connection between computational
data-mining (in particular, diffusion maps and the resulting
low-dimensional description of high-dimensional data) with
computational multiscale methods (in particular, certain equation-free
algorithms).
Our illustrative example consisted of a model gene regulatory network
known to exhibit bistable (switching) behavior in some regime of its
parameter space.
We also presented examples of {\it lifting} and {\it restriction}
protocols, that enable the passing of information between detailed
state space and reduced ``diffusion map coordinate" space.
These protocols allow us to ``intelligently" design short bursts of
appropriately initialized stochastic simulations with the detailed
model simulator.
Processing the results of these simulations {\it in diffusion coordinate
space} forms the basis for the design of subsequent numerical experiments
aimed at elucidating long-term system dynamic features (such as
equilibrium densities, effective free energy surfaces, escape times
between different wells, and their parametric dependence).
In particular, we confirmed that previously, empirically known,
observables were indeed meaningful coarse-grained coordinates.

In traditional diffusion map computations, a single scalar
(a scaled Euclidean norm) forms the basis for the identification
of good reduced coordinates (when they exist).
An important issue that arose in our example, due to the
disparate nature, value ranges and dynamics of different data vector components,
was the selection of appropriate {\it relative} scaling among data component values.
The computational approach we used was based on the data ensemble, without
any contribution from the {\it dynamical proximity} between data points
collected along the same trajectory.
We believe that incorporating such information will be very useful in
determining relative scalings among disparate data components; finding
ways to integrate such information among data ensembles collected in
different experiments, and possibly with different sampling rates will
greatly assist in this direction.

In this work, diffusion map computations were based on data collected
from a single long transient, that was considered representative of
the entire relevant portion of the (six-dimensional) phase space.
In more realistic problems such long simulations will be no longer
possible; yet local simulation bursts, observed on {\it locally valid}
diffusion map coordinates can be used to guide the efficient
exploration of phase space. 
Local smoothness in these coordinates allows us to use them in
protocols such as umbrella sampling \cite{Torrie:1974:MCF,Ryckaert:1977:NIC} to ``differentially locally
extend" effective free energy surfaces. For example, ``reverse coarse'' integration described in
\cite{Gear:2004:CPF,Frewen:2006:EFE} provides computational protocols
for microscopic/stochastic simulators to track backward in
time behavior, accelerating escape from free energy minima
and allowing identification of saddle-type coarse-grained
``transition states''.
Design of (computational) experiments for obtaining macroscopic
information is thus complemented by the design of (computational)
experiments to extend good low-dimensional data representations:
both the coarse-grained coordinates {\it and} the operations we
perform on them can be obtained through appropriately designed fine
scale simulation bursts. 

In this paper the connection between diffusion maps and
coarse-grained computation operated only in one direction:
diffusion map coordinates influenced the subsequent design of
numerical experiments.
An important current research goal is to establish the ``reverse
connection": the on-line extension/modification of diffusion map
coordinates towards sampling important, unexplored regions
of phase space.

\section*{Acknowledgements}

This work was partially supported by DARPA (TAF, RC, IGK, BN), NIH Grant
R01GM079271-01 (TCE, XW), and the Biotechnology and
Biological Sciences Research Council and Linacre College,
University of Oxford (RE).

\bibliographystyle{amsplain}
\bibliography{VFEF22}

\appendix

\newpage

\appsection{Diffusion Maps}

\label{appDMAP}

The following discussion is largely adapted from \cite{Belkin:2003:EDR,Coifman:2005:GDT}.
We
present a criterion for dimensionality reduction and show how it
leads to the diffusion map method.

Suppose we have $M$ points ${\bf X}^{(i)}\in \mathbb{R}^N$,
$i=1,\ldots,M$, and we define the matrix ${\bf W}$ by (\ref{weightmatrix}).
Given a mapping ${\bf f}:[1,\ldots,M]\to\mathbb{R}^n$, we define the
functional ${\cal L}$ by the formula
\begin{equation}
  {\cal L}({\bf f}) = \sum_{i,j} \norm{{\bf f}(i) - {\bf f}(j)}^2 W_{ij}.
    \label{L_f}
\end{equation}
We see that ${\cal L}({\bf f})$ is always nonnegative.
Moreover,
$W_{ij}$ is close to (resp. far from) one for vectors ${\bf
X}^{(i)}$ and ${\bf X}^{(j)}$ which are near (resp. far) from each
other.
For a dimensionality reduction function ${\bf f}$ to be
useful, we must make sure that nearby points ${\bf X}^{(i)},{\bf
X}^{(j)}$ in $\mathbb{R}^N$ are mapped to nearby points ${\bf
f}({\bf X}^{(i)}),{\bf f}({\bf X}^{(j)})$ in $\mathbb{R}^n$.
To find
such a mapping, one can solve the following minimization problem
\begin{equation}
  \arg\min_{{\bf f}\in\mathbb{F}} {\cal L}({\bf f})\quad\mbox{ where  } \mathbb{F} =
  \{{\bf f} : {\bf F}^T {\bf D} {\bf F}={\bf I}_n, {\bf F}^T {\bf D}{\bf 1} = {\bf 0}\}
    \label{min_L_f}
\end{equation}
where ${\bf F}$ is the $M\times n$ matrix with row vectors ${\bf
f}(i)$, ${\bf D}$ is the $M\times M$ diagonal matrix with entries $D_{ii} = \sum_j W_{ij}$
, $i=1,\ldots,M$, ${\bf I}_n$ is the $n\times n$ identity
matrix, $\bf 1$ is a vector of $M$ ones, and ${\bf 0}$ a vector of
$n$ zeros.
The first constraint removes the arbitrary scaling
factor, while the second constraint ensures that we do not map all
$M$ points ${\bf X}^{(i)}$ to the same number.
Since (\ref{L_f}) can
be rewritten as
\begin{equation}
  {\cal L}({\bf f}) = \sum_{i,j=1}^M \norm{{\bf f}(i) - {\bf f}(j)}^2 W_{ij} =
  \mbox{tr}({\bf F}^T({\bf D}-{\bf W}) {\bf F})
\end{equation}
the solution ${\bf F}$ is given by the matrix of eigenvectors
corresponding to the lowest eigenvalues of the matrix
\begin{equation}
  {\bf D}^{-1} [{\bf D} - {\bf W}] = {\bf I}_M - {\bf K}
\end{equation}
or equivalently by the largest eigenvalues of ${\bf K}$.
By the
non-negativity of the functional ${\cal L}({\bf f})$ it follows that
the eigenvalues of ${\bf I}_M-{\bf K}$ are all non-negative, or that
all eigenvalues of ${\bf K}$ are smaller than or equal to one.
The
eigenvector corresponding to the eigenvalue $\lambda_0=1$ is the
vector ${\bf 1}$. Ordering the remaining eigenvectors in decreasing
order we see that the $n$-dimensional representation of
$N$-dimensional data points, via the minimization of (\ref{min_L_f})
is the diffusion map (\ref{lowdimenrepresentation}).

We note
that our $M$ points and the matrix $W_{ij}$ can be also viewed as the
weighted full graph with $M$ vertices, where the weight associated
with an edge between points $i$ and $j$ is equal to $W_{i,j}$.
Then the previous analysis can be reformulated in terms
of standard spectral graph theory
\cite{Chung:2000:HEI,Belkin:2003:EDR}.
More precisely, it was shown in \cite{Nadler:2005:DMS2} that
this construction leads to the classical normalized
graph Laplacian for $\alpha=0$ in (\ref{weightmatrix}).
If $\alpha=1$, then the construction gives the
{\em Laplace-Beltrami} operator on the graph.
Finally, if the data are produced by a stochastic (Langevin)
equation, $\alpha=1/2$
provides a consistent method to approximate
the eigenvalues and eigenvectors of the underlying
stochastic problem.

\appsection{Simple Illustrative Examples}

\label{DMAPexamples}

\begin{figure}[htbp]
\centerline{
\psfig{file=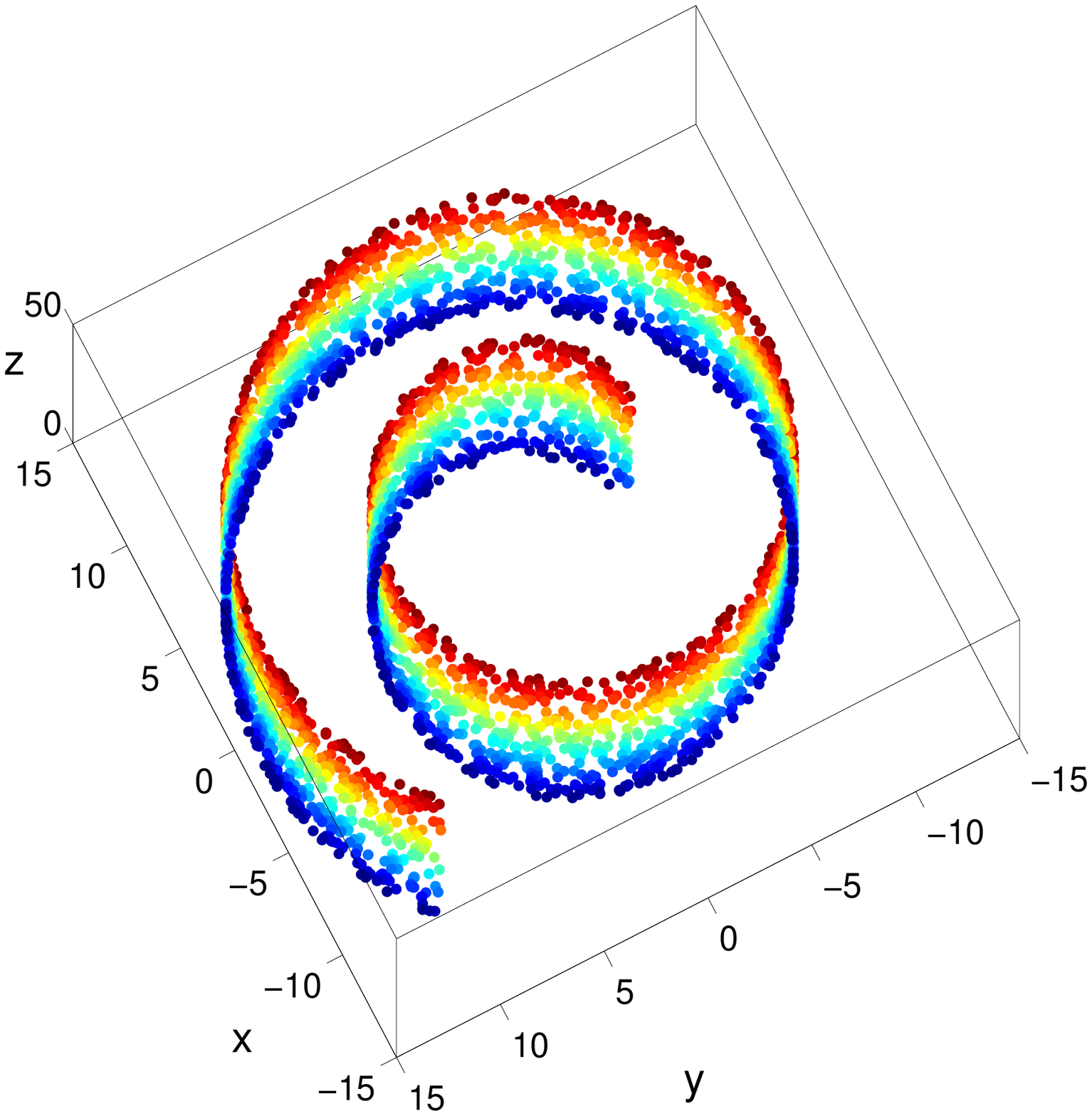,height=2.5in}
\psfig{file=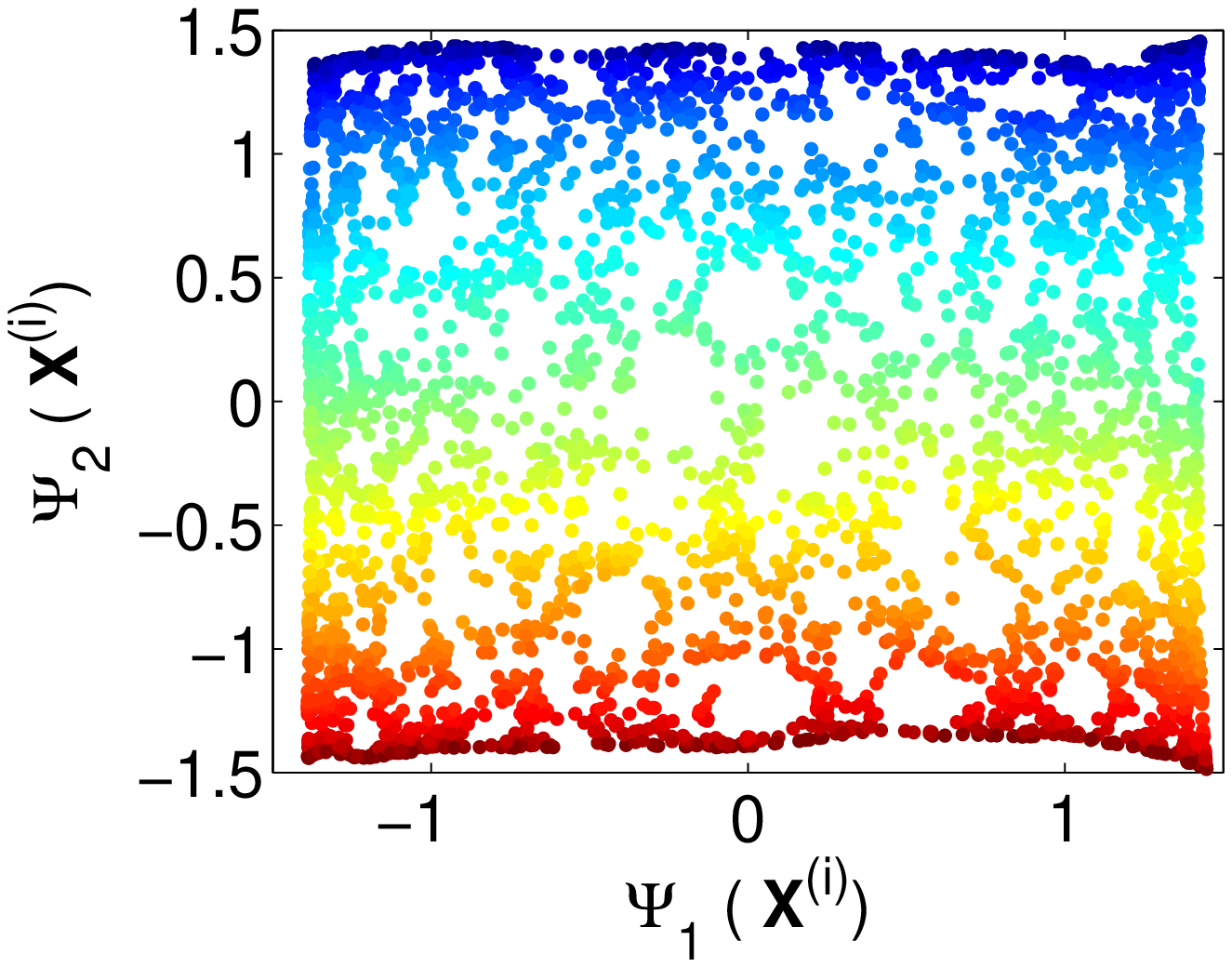,height=2.5in}
}
\caption{{\it Left panel: Swiss roll dataset in $\er^3$. Datapoints
    lie along a $2$-dimensional manifold. Datapoints are colored by 
    their $z$-coordinate
    value (ordering of datapoints passed to diffusion map routine is
    random). Right panel: plot of $\Psi_1 (\boldX^{(i)})$
 (corresponding to eigenvalue $\lambda_1$) against $\Psi_2
    (\boldX^{(i)})$ (corresponding to eigenvalue $\lambda_1$) for points in the dataset (same coloring
    scheme). The diffusion map ``unrolls'' the
    $2$-dimensional manifold. }}
\label{figSwissroll+DMAP}
\end{figure}

We include a brief illustration of the application of the diffusion map approach to the well
known 3-dimensional ``Swiss roll'' data set \cite{Tenenbaum:2000:GGF,Roweis:2000:NDR,Donoho:2003:HEL} (shown in left
panel of Figure \ref{figSwissroll+DMAP})
where datapoints lie along a 2-dimensional manifold. 
For this dataset $\boldX = \left[x,y,z\right]$; to compute the
diffusion map we use $\alpha=1$, and $\sigma=2$ in equation (\ref{weightmatrixtilda0}).
Figure
\ref{figSwissroll+DMAP} (right panel) plots these datapoints in terms of their components in the top two
significant eigenvectors ($\Psi_1 (\boldX^{(i)}), \Psi_2 (\boldX^{(i)})$) of the matrix $\boldK$ for this
dataset; it shows the ``unrolled'' 2-dimensional manifold detected by
the diffusion map algorithm.
The same result is obtained irrespective
of the ordering (or orientation) of  the dataset used to
compute the pairwise similarity matrix.

\begin{figure}
\centerline{
\psfig{file=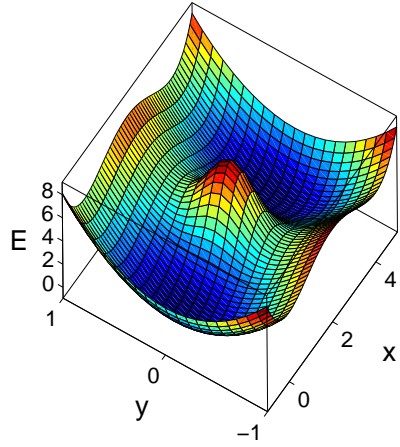,height=2.5in}
}
\caption{{\it Two-well potential $(\ref{2wellPOT})$ 
with two connecting pathways between minima.}}
\label{fig2wellPOT}
\end{figure}

\begin{figure}
\centerline{
\psfig{file=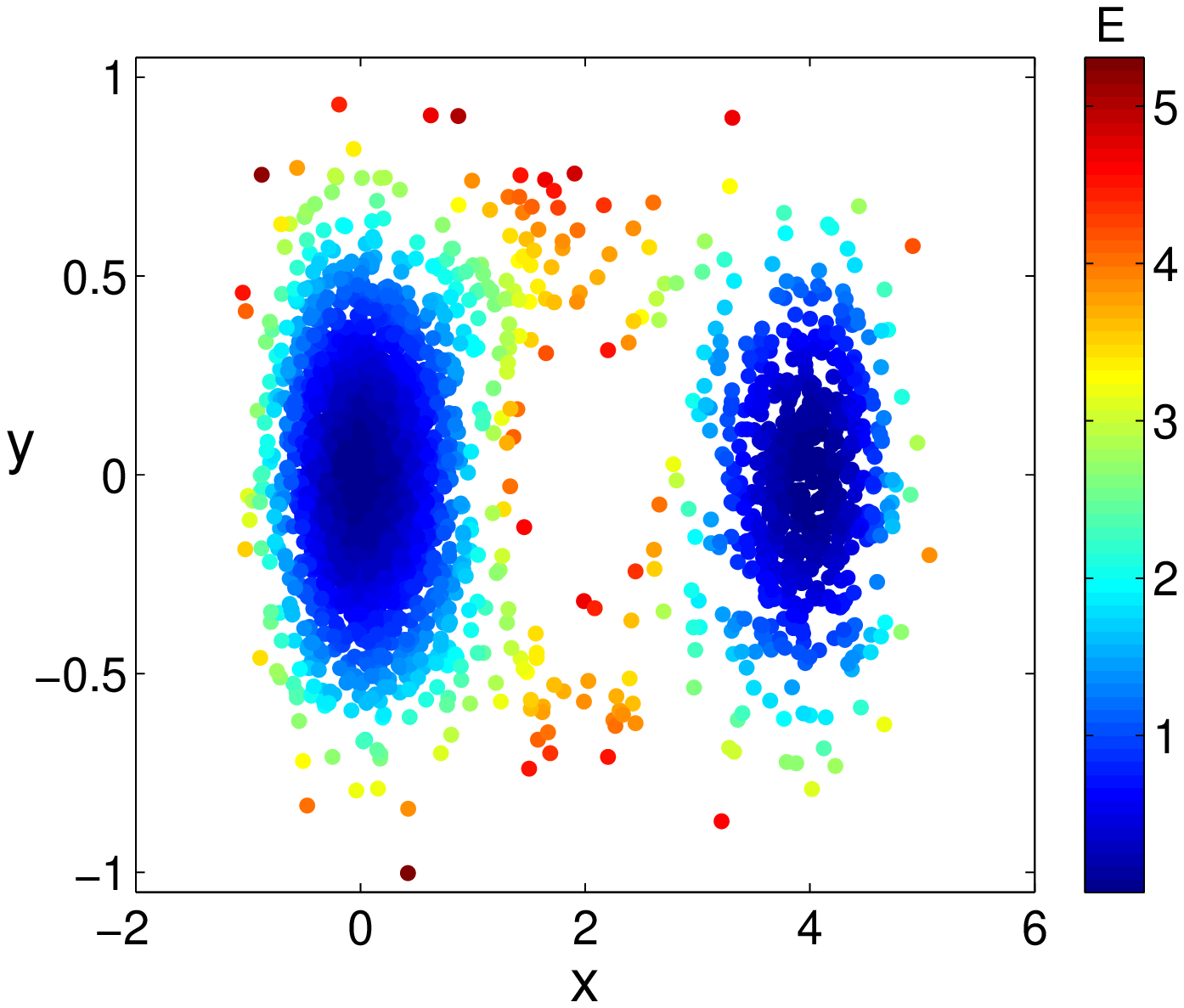,height=2.5in}
\psfig{file=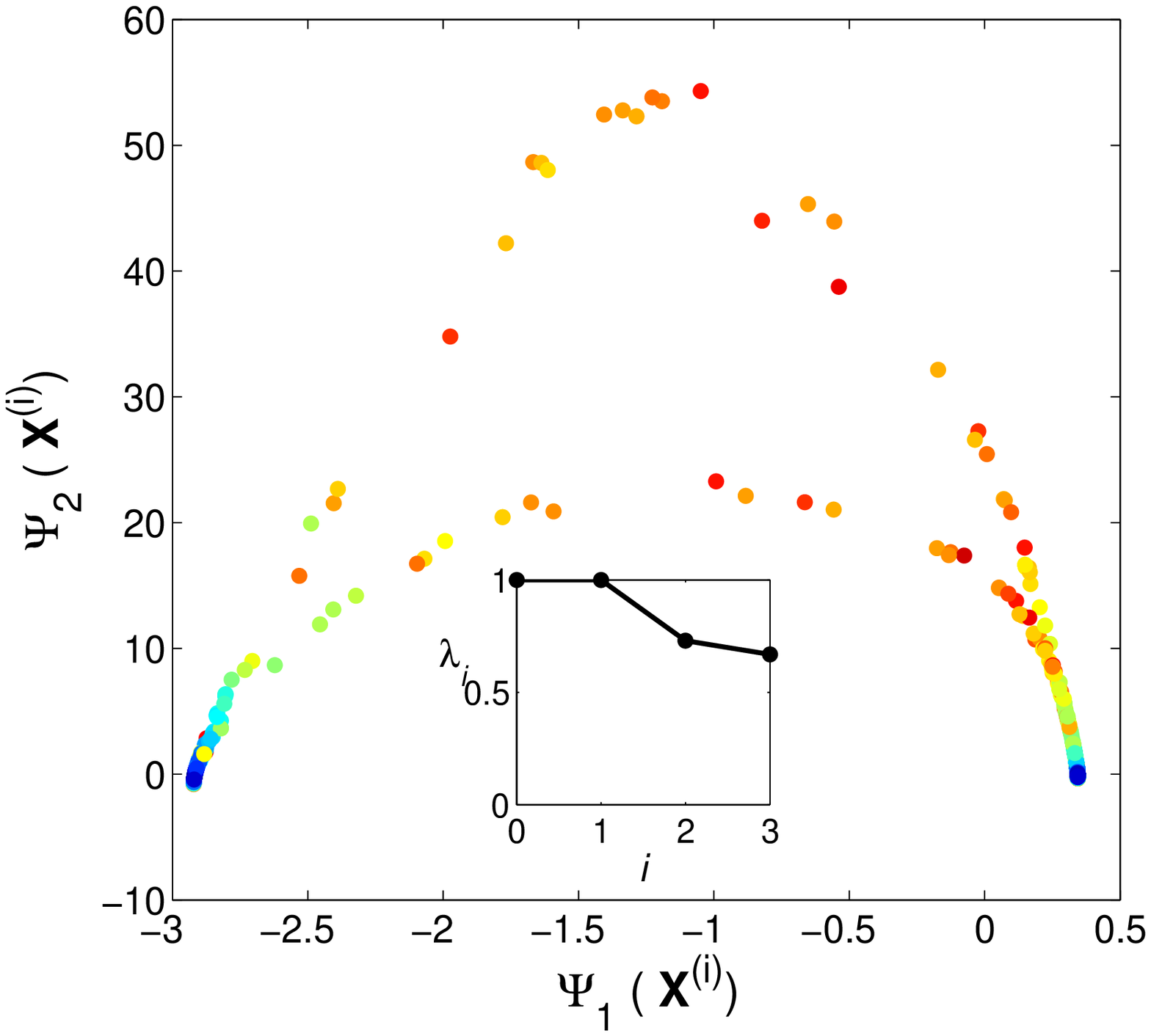,height=2.5in}
}
\caption{{\it Left panel: Subsampled dataset generated by Monte Carlo
    simulation using potential defined in equation $(\ref{2wellPOT})$
    (datapoints colored by energy according to colorbar). Right
    panel: dataset diffusion map (same coloring scheme) with top
    eigenvalues indicated in inset.}}
\label{fig2wellsubsampleDMAP}
\end{figure}

As a second illustration, Figure \ref{fig2wellPOT} shows the potential

\begin{equation}
E(x,y) =
\frac{x^4}{8}-x^3+2x^2+\frac{y^4}{5}+6\exp(-2(x-2)^2-10y^2)
\label{2wellPOT}
\end{equation}
which has two minima connected by two paths.
A subsampling of the
dataset generated by Monte Carlo simulation using this potential is
shown in Figure \ref{fig2wellsubsampleDMAP} (left panel) with the corresponding
diffusion map shown in the right panel of the figure. For this dataset $\boldX = \left[x,y\right]$; to compute the
diffusion map we use $\alpha=0$, and $\sigma=0.5$ in equation (\ref{weightmatrixtilda0}).
Figure
\ref{fig2wellsubsampleDMAP} (right panel) shows that points close to the
bottom of the wells are
mapped to tight clusters in the diffusion map, with a clear distinction
between datapoints on each of the two transition pathways between the minima.

\end{document}